# Improved Methods for Fluorescence Background Subtraction from Raman Spectra

P.J. Cadusch[1], M.M. Hlaing[2], S.A. Wade[2], S.L. McArthur[2], P.R. Stoddart[2]

[1]Faculty of Engineering and Industrial Science, Swinburne University of Technology, John Street, Hawthorn, Victoria 3122, Australia

[2]Biotactical Engineering, Industrial Research Institute Swinburne, Swinburne University of Technology, John Street, Hawthorn, Victoria 3122, Australia

**Abstract**

Raman spectroscopy has attracted interest as a non-invasive optical technique to study the composition and structure of a wide range of materials at the microscopic level. The intrinsic fluorescence background can be orders of magnitude stronger than the Raman scattering and so background removal is one of the foremost challenges for quantitative analysis of Raman spectra in many samples. A range of methods anchored in instrumental and computational programming approaches have been proposed for removing fluorescence background signals. An enhanced adaptive weighting scheme for automated fluorescence removal is reported, applicable to both polynomial fitting and penalised least squares approaches. Analysis of the background fitting results for ensembles of simulated spectra suggests that the method is robust and reliable, and can significantly improve the background fit over the range of signal, shot-noise and background parameters tested, while reducing the subjective nature of the process. The method was also illustrated by application to experimental data generated from aqueous solutions of bulk protein fibrinogen mixed with dextran.

## 1. Introduction

Raman spectroscopy has been widely applied in materials characterization. For example it has attracted attention in biology as a non-destructive tool for characterizing the chemical properties of cells, as well as subtle molecular and biochemical changes within cells.[1] In biomedical applications it has been used for rapid microbial identification, characterization of biomolecular changes associated with disease transformation and studying the process of protein adsorption on implant materials.[2,3] A significant challenge for many applications of Raman spectroscopy is that the spectra are often accompanied by noise superimposed on a broad background. This background is generally dominated by intrinsic fluorescence from the sample.[4] Consequently, the fluorescence background has to be removed in order to perform further quantitative analysis on the Raman spectra, including multivariate analysis.

In order to remove fluorescence background from measured Raman signals, approaches based on instrumental, experimental and computational methods have been widely applied.[5] Instrumental approaches to minimise the fluorescence background, such as excitation wavelength shifting, time-gating and photo-bleaching, require hardware modifications in the spectroscopic system.[6-8] The excitation wavelength shifting technique requires two closely spaced excitation wavelengths to achieve two spectra and further processing to fit the Raman spectrum. Although it requires some system modification, it has been reported to eliminate both the fluorescence background and systematic noise from the spectra.[9] There are a number of reported attempts to develop a time-gating system to solve the problem of low signal-to-noise ratio (SNR) spectra, but there are difficulties in system modification to achieve low peak power pulses with high gating efficiency at a safe threshold for biological samples.[10,11] Photo-bleaching of samples has been proposed to reduce the broad fluorescence background, but the relative heights of Raman peaks obtained from the sample are progressively altered as a consequence of the irradiation and the removal of fluorescence background from the samples may be inadequate.[12,13] Experimental approaches such as selection of substrates (calcium fluoride or zinc selenide) and improved sample preparation have been proposed in order to increase the data quality by minimizing the fluorescence background.[14] However, substantial background remains in the data due to the interaction between the light source and the intrinsic fluorescence in many samples.

As a result of these challenges, computational methods have become the standard way to correct for contributions from fluorescence in the background. These require no system modifications and impose no limitations on sample preparation. Among these mathematical techniques, first- and second-order derivatives, frequency-domain filtering, polynomial fitting and wavelet transformation methods have been proposed as useful tools for background removal in certain situations.[5] The accuracy of the first- and second-order derivative methods is reliant on peak selection. Due to the difficulty of peak picking, missing some peaks could result in aspects of the Raman spectrum being placed in the baseline, resulting in a poor baseline estimate. The first- and second-order derivative methods can severely diminish and distort Raman spectral features unless there are complex mathematical fitting algorithms to reproduce a traditional spectral form.[15,16] Fast-Fourier transform filtering (FFT) is one of the frequency-domain filtering techniques and also requires the separation of the frequency components of the Raman spectrum from those of the background and noise.[17] Polynomial fitting has become the most popular fluorescence removal technique for a wide range of applications. However, manual polynomial fitting relies on user intervention for selection of locations where the curves are to be fitted in the data. Although automatic polynomial fitting methods have been proposed to remove the need for manual curve-fitting, their use can be limited in high noise circumstances.[18-20] Wavelet transform methods can also be used to automate the curve fitting, but difficulties in the selection of suitable wavelet thresholds and the proper level of resolution to represent the baseline may affect the background removal results.[21]

Recently, a background-correction algorithm for Raman spectra has been developed using wavelet peak detection, wavelet derivative calculation for peak width estimation, and penalized least squares background fitting.[22] This approach adaptively separates the measured data samples into peak and non-peak (background) values by setting the least-squares weights to one for background and zero for peak regions. The application of these binary valued weights may cause some sudden changes in gradient that appear questionable in the context of a Raman background subtraction. In this paper, we propose enhanced automated algorithms for fluorescence removal based on a combination of adaptive weighting with the penalized least squares estimation and also with polynomial estimation. Experimental results show that the proposed method can automatically identify background regions and that the results are comparable with or superior to previously reported methods for fluorescence background subtraction.

## 2. Materials and Methods

### 2.1. Background removal using adaptive-weight penalised least squares (APLS)

The fluorescent background is relatively smooth on the scale of the Raman peaks from most samples. In order to separate the Raman peaks from the fluorescent background, we follow the general approach of background correction for Raman spectra based on penalized least squares (PLS) background fitting.[22] In that approach, the background is based on the values $\{b_n; n = 1,2,3,\cdots,N\}$ which minimize the spline-type cost function applied to the observed values $\xi_n$. To achieve a trade-off between fidelity to the data and roughness of the derived background, the cost function can be described as follows:

$$C_m(\mathbf{b},\gamma) = \sum_{n=1}^{N} w_n (b_n - \xi_n)^2 + \gamma \sum_{n=1}^{N-m} \left[ \left( \mathbf{D}^{(m)} \mathbf{b} \right)_n \right]^2, \tag{1}$$

where $\{w_n : n = 1,2,\cdots,N\}$ are the weighting factors that are to be adaptively determined, $N$ is the number of wavenumbers at which the spectrum is measured and $\mathbf{D}^{(m)}\mathbf{b}$ is an $m^{th}$ difference operator, (we consider only the first- and second-order versions which are $\left(\mathbf{D}^{(1)}\mathbf{b}\right)_n = b_{n+1} - b_n$ and $\left(\mathbf{D}^{(2)}\mathbf{b}\right)_n = b_{n+2} - 2b_{n+1} + b_n$). The smoothing parameter, $\gamma$, is an adjustable parameter of the spline smoothness. The larger the smoothing parameter, the stronger the impact on roughness of the derived background and the smoother ("stiffer") the spline curves. In the continuous version (with integrals replacing sums over sampled values) when $m = 1$, the fit is continuous and the gradient is square-integrable. When $m = 2$, the gradient is also continuous and the second derivative is square-integrable. The discrete fits tend to display similar properties, so that the $m = 1$ case produces continuous fits with occasional sudden changes in gradient whereas the $m = 2$ case produces smoother fits without the gradient discontinuities.

We have developed an approach that simplifies both the adaptive region specification and the negative-value avoidance steps of Zhang *et al* [22] by treating the entire signal point-wise as a combination of two Poisson processes:

$$\xi_n = r_n + b_n, \tag{2}$$

where $r_n$ is the Raman count and $b_n$ the fluorescent background count (plus dark counts etc.). Assuming that a crude estimate of the background mean, $\mathbf{b} = [b_1 \ b_2 \ \cdots \ b_N]^T$, is available at the first step, we define the weighting factors, $\mathbf{w} = [w_1 \ w_2 \ \cdots \ w_N]^T$, used in the PLS fit to be the probabilities that the observed values ($\xi_n$) or greater, would arise by chance from the background alone given the current estimate of the background mean level. That is,

$$w_n = \Pr(\xi \geq \xi_n \mid b_n). \tag{3}$$

The process is repeated using the new weights, $\mathbf{w}$, to determine new estimates of $\mathbf{b}$ which minimise the cost-function, which are then used in turn to determine new weights and so on, until the weights settle down.

It should be stressed that, apart from an initial guess at the background (which is not critical; the mean of the observed signal appears to be an appropriate choice) the procedure requires only the setting of the smoothing parameter $\gamma$, but this needs to be done intelligently, based on the expected type of background. Too small a value of $\gamma$ results in most points being treated as background, thus losing genuine peaks, whereas too large a value leads to most points being taken as non-background, thus degrading the background estimation. The value also depends on the number of points sampled and to some extent on the scaling of the data. In addition, the required value for first- and second-order difference schemes can differ by several orders of magnitude (2nd order fits with $\gamma \approx 25000$ are similar to 1st order fits with $\gamma \approx 10$). Therefore some experimentation is needed to find a reasonable value of the smoothing parameter. Efforts are continuing to find criteria for setting the smoothing automatically based on the following observations. Firstly, a successful estimation should result in reduced correlation between the estimated background and the resulting background subtracted spectrum. Secondly, the confidence one has in an extracted background should decrease as the number of points used for the estimation decreases. To date the results appear no better than those of an educated interactive setting of the smoothing based on a typical spectrum and will not be further discussed in this paper.

## 2.2. Improved background removal using doubly weighted spline

The rationale behind the cost function can be extended to relax the effect of global stiffness. When the data contain overlapping peaks, the region common to the peaks can appear to the fitting routine to be part of the background. Using a stiffer spline can ameliorate this at the expense of less flexible fitting in the true background regions or in adjacent non-background regions. If, however, the stiffness of the spline is defined adaptively, so that it is larger in non-background regions than in background regions, then the fitted background can better respond to variations in the general properties of the data. The simplest way to do this is to weight the roughness term in the cost function so that it is less important in regions considered background and more important in non-background regions. The simplest weighting scheme is to weight the roughness term with a linear interpolation between the maximum stiffness when $w_n = 0$ (i.e. in peak regions) and the minimum stiffness when $w_n = 1$ (i.e. in the background regions). This can be described as follows:

$$C_m(\mathbf{b},\gamma) = \sum_{n=1}^{N} w_n (b_n - \xi_n)^2 + \sum_{n=1}^{N-m} [\gamma_{\min} w_n + \gamma_{\max}(1-w_n)] [(\mathbf{D}^{(m)}\mathbf{b})_n]^2. \quad (4)$$

More conveniently, it can be described by the following equation:

$$C_m(\mathbf{b},\gamma) = \sum_{n=1}^{N} w_n (b_n - \xi_n)^2 + \gamma_{\max} \sum_{n=1}^{N-m} (1-\eta w_n) [(\mathbf{D}^{(m)}\mathbf{b})_n]^2, \quad (5)$$

where $\eta = (\gamma_{\max} - \gamma_{\min})/\gamma_{\max}$. For $\eta = 1$, the roughness term is ignored in a pure background region and takes its maximum value in a pure peak region. For $\eta = 0$, the cost function reduces to the single weighted version ($W^1$). Values in the vicinity of 0.9 appear to perform reasonably well, but this is somewhat data dependent. Values outside the range [0, 1] are not meaningful and should be avoided. This is referred to as the double weighted spline ($W^2$).

## 2.3. Improved polynomial background removal (APoly)

The adaptive weighting scheme described by equations (2) and (3) above, can also be applied to conventional polynomial background estimation with the cost function:

$$C(\mathbf{p}, M) = \sum_{n=1}^{N} w_n \left( \sum_{m=0}^{M} p_m k_n^m - \xi_n \right)^2 \qquad (6)$$

The advantage here is that, once the order (*M*) of the polynomial has been decided, there is no need to find a suitable value for a tuning parameter such as $\gamma$. The disadvantage is that, as with most un-regularized polynomial fitting, the problem may become ill-conditioned as the order of the polynomial is increased.

### 2.4. Simulated data

In order to assess the accuracy of the proposed algorithm, ensembles of simulated Raman data consisting of a number of Gaussian peaks (typically up to 20 peaks) and a variable background were used for testing purposes. Each ensemble consisted of 100 spectra and in each spectrum of a given ensemble the locations, amplitudes and widths of the peaks were chosen randomly from uniform distributions over ranges typical of those seen in real Raman data. The range of parameter values used in constructing the ensembles is shown in Table 1. The backgrounds were modelled as a sum of three sinusoids, for which the amplitudes and phases were chosen randomly to mimic typical observed fluorescent backgrounds. The periods of the background sinusoids were fixed (see Table 1) for the ensembles used here and the average peak to background ratio was set at different levels for different ensembles by adjusting the upper limit of the range of peak amplitudes. These signals were sampled on a wavenumber grid that matches that of the experimental Raman data.

Given the low background noise levels in cooled CCD arrays, the signal levels in modern Raman spectrometers are generally shot noise limited. Therefore the simulated test signal was constructed by replacing each sampled value calculated as above by a Poisson random variate using the sampled value as the mean. The simulated backgrounds were similarly constructed. The simulated experimental record at a given wavenumber is thus the sum of the corresponding signal and background variates. The baseline fitting routines were then tested by comparing the recovered background with the simulated known background.

To assess the effect of shot noise on the quality of the fits, the simulated spectra were rescaled to either increase or decrease the expected shot-noise. Since we are interested in the quality of the recovered background we have used the background signal to shot noise ratio, rather than the full signal to noise ratio, as a measure of the signal condition. A confounding effect, which may be obscured when using the full signal to noise ratio, is the fact that the larger the relative size of the peaks, the harder it is to derive a good background. This is due to the tendency of the overlapping peaks to resemble background features, especially as the number of peaks in a given wavenumber range is increased. For the structured background ensembles described above, the background to shot-noise ratio (BNR) is defined as:

$$\text{BNR} = \frac{\overline{b(k)}}{\sqrt{\overline{r(k)} + \overline{b(k)}}} \qquad (7)$$

where $\overline{b(k)}$ is the mean of the simulated background signal over the range of wavenumbers used, and $\overline{r(k)}$ is the mean of the simulated (background-free) Raman spectrum.

The quality of the fit between the extracted background and the known background depends on the extent to which the variation of the background resembles that of the peak data. For the penalised least squares methods there is usually an optimum value for $\gamma_{max}$ which, in the double weighted case, also varies with the value chosen for $\eta$. To get a general idea of the likely errors in background fitting, ensembles of 100 simulated spectra were generated as described before and were analysed for each condition tested (varying number of peaks, peak to background ratio and signal to noise ratio) using the range of parameters shown in Table 1. For convenience in presenting the results a normalised error ratio was defined to allow comparison of background fits for varying parameter settings:

$$\rho = \frac{\sqrt{\overline{(b(k)-\hat{b}(k))^2}} - \sqrt{\overline{b(k)}}}{\sqrt{\overline{(b(k)-\overline{b(k)})^2}}}$$

$$= \frac{\text{RMS residual background - expected residual background}}{\text{RMS background variation}} \quad (8)$$

where $b(k)$ is the known background, $\hat{b}(k)$ is the estimate of the background and the over-bar denotes an average over the wavenumbers $k$. It should be noted that the offset from the expected background error is used since the data are intrinsically stochastic. Moreover, removing an ideal mean background will always leave some residual error. Therefore, the measure (normalised error ratio) aims to gauge the improvement obtained in estimating the background rather than the absolute error. In theory this measure could be negative, indicating over-fitting of the background noise in low peak to background data at low SNRs, but this has rarely been seen in the data analysed here. Thereafter box plots of the error ratio were plotted for $\gamma$ values in the vicinity of the optimum $\gamma$; for various peak to background ratios; for various values of the intrinsic background to shot-noise ratio and for varying numbers of Raman peaks. Box plots were chosen over a simple mean and standard deviation since the distribution of error ratios is markedly asymmetric. The numerical values of the median and upper and lower quartile boundaries are also tabulated for the ensembles tested (Tables 2 and 3).

Extracted backgrounds were calculated using five different methods:

1. The Modified Polyfit method of Lieber and Mahadevan-Jansen [23] which uses least squares fitting of a polynomial background with (in effect) adaptive elimination of peak regions from the fit (ModPoly).

2. The Improved Modified Polyfit method of Zhao et al [19] which is similar to ModPoly but improves the peak removal scheme to allow for statistical variations in measured quantities and includes an automated iteration cut-off (IModPoly).

3. The probability-based adaptively weighted polynomial fit method described above (APoly).

4. The method of Zhang et al [22]: a penalised least squares method which differs from APLS in the way in which the adaptive weights are set (wavelet peak detection with hard peak / background segmentation) (WPLS).

5. The probability-based adaptive weight penalised least-squares method described above (APLS).

The main difference between the methods proposed here (APoly and APLS) and the other similar methods (ModPoly, IModPoly and WPLS) is that the weighting schemes in the existing methods, which consist of hard background / foreground segmentation schemes of increasing complexity, are replaced by a single, simple weighting scheme based on the statistical properties of the signal.

Matlab™ code for the ModPoly, IModPoly, APoly and APLS tests was custom written, and the R code of Zhang *et al.* for Baseline Wavelet version 4.0.1 [24] was used for the WPLS method.

The polynomial fitting routines (1), (2) and (3) above are, by their nature, better adapted to relatively simple (typically monotonic) background variation than to the more structured backgrounds of the simulated ensembles described above. To allow for this, other sets of common ensembles based on randomised versions of the simulated spectra used in Zhao *et al.* [19] were constructed. The members of these ensembles have the same peaks as in Zhao *et al.* but with their locations randomly shuffled. The background was also varied by randomly changing the polynomial coefficients [25] of Zhao *et al.* by amounts which result in a maximum of roughly 10% variation in the contribution of each of the terms in the polynomial. This approach maintained the essential characteristics of the spectra and background while allowing for a reasonable degree of variability. Also, in order to allow comparison with the present methods (APoly and APLS), which expect data values consistent with a Poisson distribution rather than the deterministic spectra used in Zhao *et al.*, the peak and background values were appropriately scaled and then used as means for Poisson variates, as in the first set of simulated ensembles. Similarly, the intrinsic BNR of the spectra could be varied by scaling the simulated signals while maintaining a constant peak to background ratio.

**2.5. Experimental data**

To demonstrate the application of the fluorescence background correction algorithms, Raman spectra were obtained from a mixture of dextran and fibrinogen. The intention of this mixture was to model a combination of biomolecules (i.e. polysaccharides and proteins) found in typical biological samples. Briefly, the suspension of dextran (Fluka, 24 µM) and fibrinogen fraction I from bovine plasma (Sigma, 29.4 µM) dissolved in MiliQ water (pH 7.4) were mixed together to get a final molar ratio of 1:8. Droplets of 10 µl of the sample mixture were air dried on a quartz slide for Raman analysis.

A Renishaw InVia Raman spectrometer equipped with a Leica microscope, deep depletion charge-coupled device detector, 2400 lines per mm grating, holographic notch filter and ~ 7 mW of 514 nm radiation from an argon-ion laser was used for acquiring spectra from the sample. The system was calibrated and monitored using a silicon reference (520.5 $cm^{-1}$) before the measurements. For each measurement, the sample was brought into focus using a 50× microscope objective (NA = 0.75 in air). The accumulation time for one spectrum was 10 s and three accumulations were collected for a single measurement on each sample area. The spectra were then averaged over three different sample areas. These Raman spectra were used to illustrate the proposed background removal algorithms using experimental data. A comparative study was performed using the five different background correction methods described above.

## 3. Results and Discussion

### 3.1. Simulated data

*3.1.1. Effect of APLS order and weighting*

A typical simulated spectrum is shown in Fig. 1(a), together with the artificial background. The background spectra produced by four different implementations of the adaptive algorithm APLS are shown in Fig. 1(b): first-order, single-weighted (referred to here as $O^1W^1$), first-order, double-weighted ($O^1W^2$), second-order, single-weighted ($O^2W^1$), and second-order, double-weighted ($O^2W^2$). For these background removal processes, where appropriate, $\eta$ has been set to 0.5 and the $\gamma_{max}$ value has been optimised. This example shows that the single-weighted form tends to perform as well or better than the double-weighted form for smooth backgrounds although this can vary with the exact details of the data waveform. This conclusion is supported by the results of the ensemble tests shown in Fig 2(a). In practice, the double-weighted fit allows the stiffness to increase markedly (by factors on the order of 5 to 10) while maintaining a good fit in rapidly changing background regions. In some situations, the double weighting may be preferred although the single-weighted scheme performs as well or better in most cases with reasonably smooth backgrounds. In terms of spline order, the second-order fit performs relatively better than the first-order fit, and the single- and double-weighted versions perform equally well for the second-order spline fit (Fig. 1(b), iii and iv and Fig 2(a)). In the remainder of this paper the second-order, single-weighted background removal scheme ($O^2W^1$) is chosen for proof-of-principle of the proposed algorithm with both simulated and experimental data. Fig. 2(b) also suggests that the performance of the routines at slightly sub-optimal values of $\gamma$ should be sufficient to allow interactive setting of the smoothing parameter without serious degradation. All of the box-plots are based on the normalised error ratio calculated using Equation 8 for ensembles of 100 simulated spectra for each condition, with parameters as defined in Table 1.

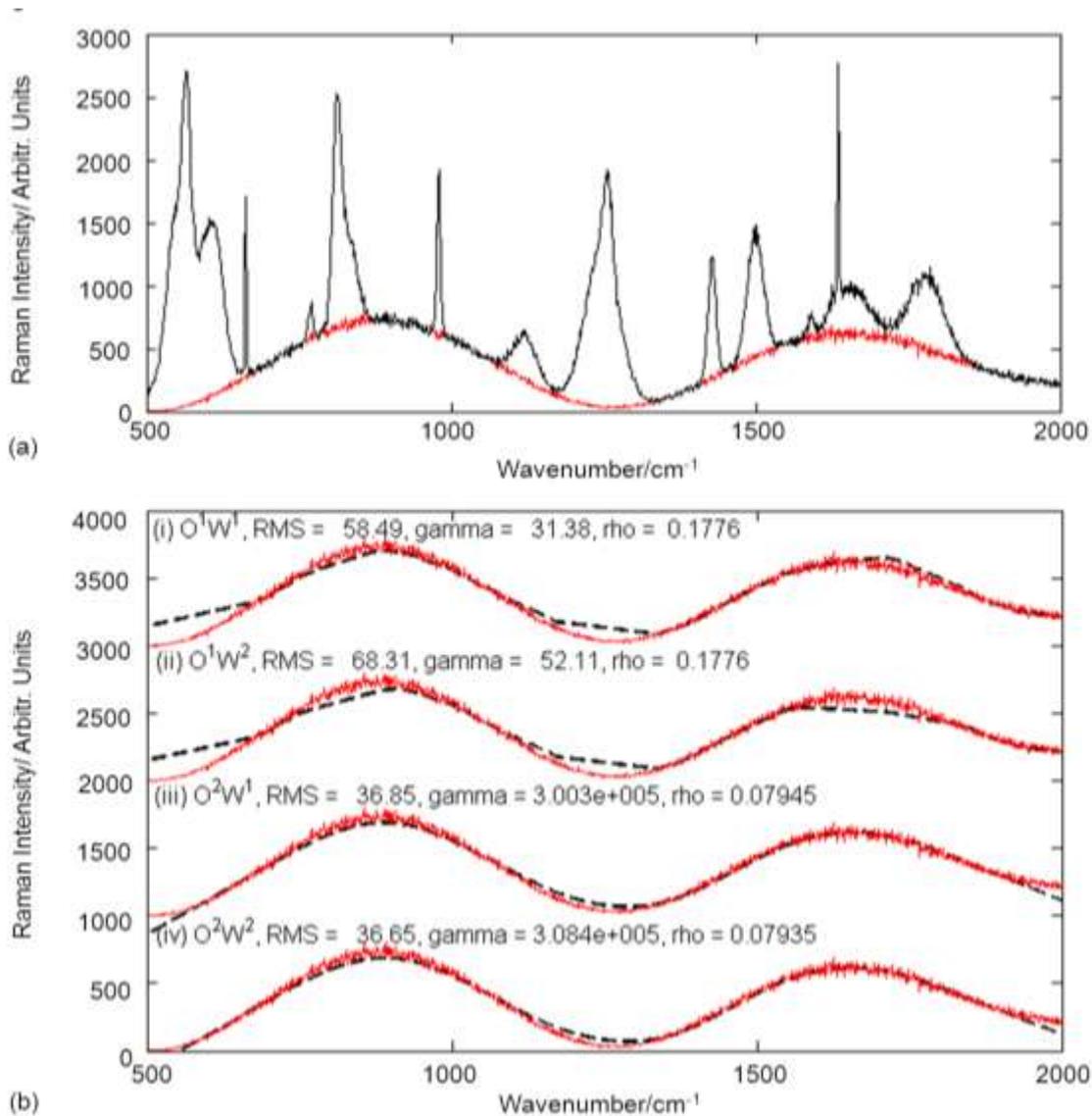

Figure 1. Simulated spectra: (a) test signal of a typical data set (20 peaks, peak to background ratio 5:1) with a simulated background, and (b) comparison between the simulated background and the recovered background for different implementations of the algorithm; (i) first-order, single-weighted ($O^1W^1$); (ii) first-order, double-weighted, $\eta = 0.5$, ($O^1W^2$); (iii) second-order, single-weighted ($O^2W^1$); (iv) second-order, double-weighted, $\eta = 0.5$, ($O^2W^2$). The solid line (red online) is the known background and dashed black line is the recovered background.

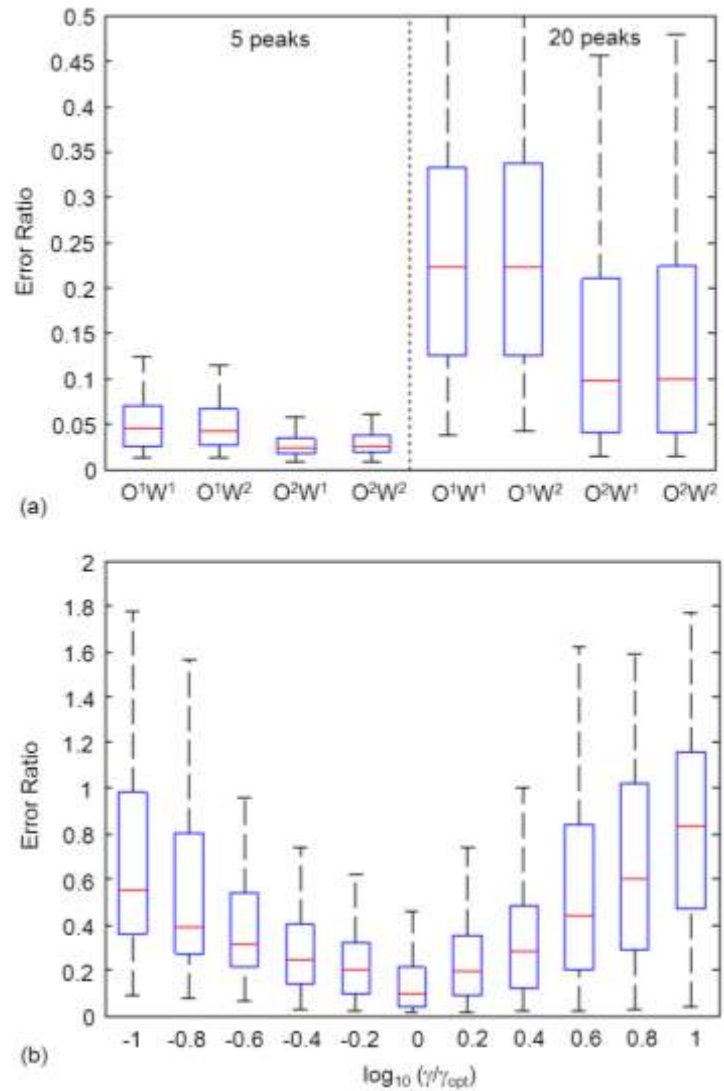

Figure 2. (a) Box plots of the error ratios for the 5 and 20-peak, high peak-to-background ratio, medium BNR simulated data sets with background estimation using the 4 variants of the APLS method. (The horizontal lines in the boxes (red on-line) are the medians, the boxes range from the 25$^{th}$ to the 75$^{th}$ percentile and the bars extend over the range of the relevant measured data set excluding outliers.) The smoothing parameters (for $\eta = 0.5$) have been optimised using the known background arrays. For the smooth background variation used here the single and double weighted versions perform equally well. (b) Box plots for a range of smoothing factors in the vicinity of the optimum values for the 20-peak common data set with 5:1 peak to background ratio using the APLS method. Distributions for smaller peak to background ratios are similar in trend but smaller in absolute value.

Table 1. Parameters for generating the simulated spectra of peaks plus background.

| Ensemble Parameters | Values |
| --- | --- |
| Background component maximum amplitude | 400 |
| Background component cycles | [1, 1.7, 2.1] |
| Number of signal (Raman) peaks | 5, 10, 12, 20 |
| Maximum Raman peak amplitude | 400, 2000 |
| **Fitting Parameters** | **Values** |
| $\gamma_{max}$ optimisation range (WPLS and APLS) | $0 - 2 \times 10^6$ |
| $\eta$ (APLS $O^1W^2$ and $O^2W^2$) | 0.5 |
| Maximum number of iterations: | |
|     APLS | 10 |
|     ModPoly, IModPoly and APoly | 250 |
| Polynomial order (ModPoly, IModPoly and APoly) | 7 |

### 3.1.2. Effect of the number of peaks

The results shown in Figs. 1 and 2 also illustrate the difficulty that arises in the presence of overlapping peaks. The simulated spectrum includes overlapping peaks in the vicinity of 606 cm$^{-1}$. Without additional *a priori* knowledge, the APLS algorithm is less effective in distinguishing between peaks and background in this region, as shown by the relatively large residual error consistently seen around 600 cm$^{-1}$ in Fig. 1(b).

This limitation is confirmed by the ensemble tests, where it was found that, for all methods tested (ModPoly, IModPoly, APoly, WPLS and APLS), the median error ratio is smaller for a smaller number of simulated peaks (Figs. 2(a) and 3(a) and Table 2). It should be noted, however, that the error is dominated by a few effects, such as overlapping peaks and peaks at or near the end of the wave number scan. From an experimental point of view, the latter issue may be avoided by selecting spectral sweeps without peaks at the extremes of the wavenumber range.

For the structured (sum of sines) background the APLS method compared favourably with all of the other methods for varying number of peaks, with the median error smaller than those for the other methods, typically by factors of 4 to 5. Though the improvement is less pronounced, APoly also performed better than ModPoly, IModPoly and WPLS for 10 and 20 peak ensembles, typically by factors on the order of 2 (Fig 3a and Table 2).

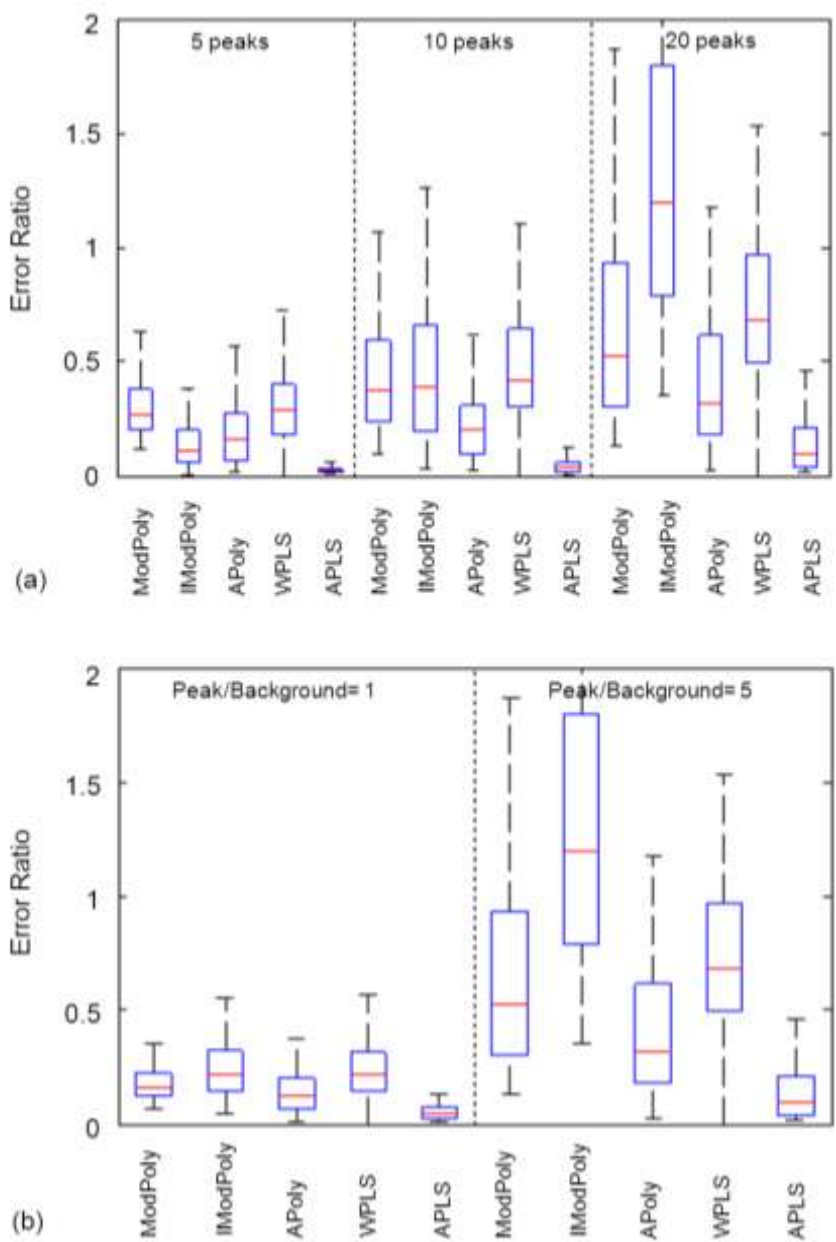

Figure 3. Box plots (defined as in Fig. 2) of the error ratio (a) for the 5, 10 and 20-peak, high peak-to background ratio, medium BNR simulated data sets. Five related methods have been tested: ModPoly, IModPoly, APoly, WPLS and APLS. The smoothing parameters for the APLS and WPLS methods have been optimised using the known background; the default first order differences have been used for WPLS (the second order differences have been tried with no significant change noted), the $O^2W^1$ variant of APLS has been used; the iteration limits for the ModPoly and IModPoly methods have been set at 250, close to the upper limit of typical values suggested by their developers, and the polynomial order is set to 7. (The recommended range of polynomial orders is typically 4 to 6 in practice, the value chosen is close to the optimum value, based on the known background for the test data used here.) (b) Box plots of the error ratios for low (1:1) and high (5:1) peak to background ratios of 20-peak simulated data ensembles, all other settings as above.

Table 2. Median error ratios and quartile boundaries for background estimation using five different methods for a range of numbers of Raman peaks and peak to background ratios.

| Signal to Background Ratio (SBR) | Number of Peaks | Method | Normalised Error Ratio | | |
|---|---|---|---|---|---|
| | | | Lower Quartile | Median | Upper Quartile |
| 5 | 5 | ModPoly | 0.200 | 0.268 | 0.382 |
| | | IModPoly | 0.060 | 0.110 | 0.202 |
| | | APoly | 0.070 | 0.161 | 0.271 |
| | | WPLS | 0.180 | 0.285 | 0.402 |
| | | APLS | 0.018 | 0.024 | 0.034 |
| 5 | 10 | ModPoly | 0.241 | 0.376 | 0.597 |
| | | IModPoly | 0.198 | 0.387 | 0.662 |
| | | APoly | 0.099 | 0.201 | 0.311 |
| | | WPLS | 0.301 | 0.416 | 0.680 |
| | | APLS | 0.021 | 0.037 | 0.041 |
| 5 | 20 | ModPoly | 0.303 | 0.525 | 0.934 |
| | | IModPoly | 0.784 | 1.199 | 1.794 |
| | | APoly | 0.181 | 0.318 | 0.613 |
| | | WPLS | 0.492 | 0.680 | 0.973 |
| | | APLS | 0.041 | 0.098 | 0.211 |
| 1 | 20 | ModPoly | 0.127 | 0.163 | 0.221 |
| | | IModPoly | 0.143 | 0.215 | 0.326 |
| | | APoly | 0.071 | 0.121 | 0.203 |
| | | WPLS | 0.146 | 0.218 | 0.314 |
| | | APLS | 0.023 | 0.044 | 0.072 |

*3.1.3. Effect of peak to background ratio*

The error ratios calculated for ensembles with different ratios of peak height to average background show an interesting trend, in that all methods performed better at the lower peak to background ratio (Fig. 3(b) and Table 2). It appears that the broader footprint of the larger peaks may trick the routines into treating the regions containing the peaks as likely candidates for inclusion in the background region. The improved performance of APLS and APoly observed for high peak to background ratios was also seen at low ratios, with factors of around 4 improvement for APLS and factors in the vicinity of 1.5 for APoly, as compared to ModPoly, IModPoly and WPLS.

*3.1.4. Effect of background to shot-noise ratio (BNR)*

For sweeps containing 12 peaks with a high peak to background ratio (5:1), the median error ratios and quartile boundaries shown in Fig. 4 and Table 3 are not significantly affected by scaling to produce BNRs of 5, 15 and 46 (13.8, 23.5 and 33.3 dB respectively). The relative improvements in error ratio for APLS and APoly compared to the other three methods seen above were also observed at different values of the BNR (defined by Equation (7)).

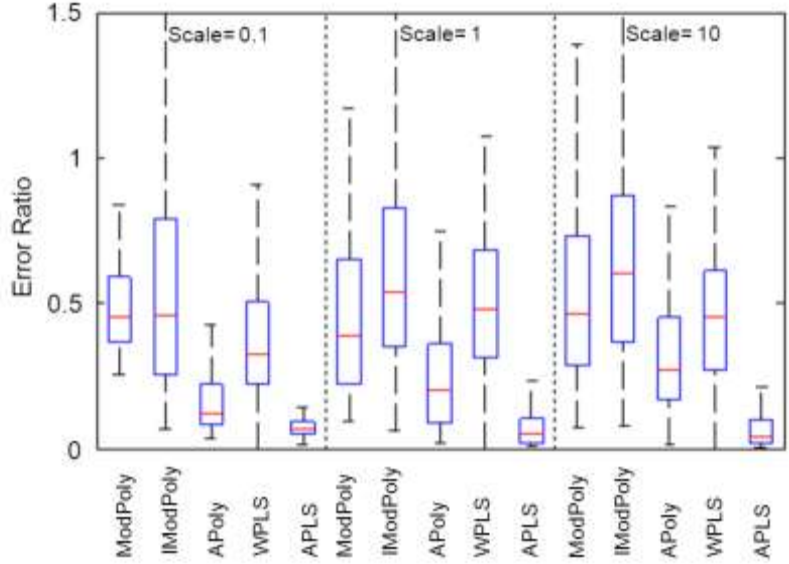

Figure 4. Error ratio of 12-peak, 5:1 peak to background ratio, structured-background data for the five methods for three increasing scale factors (equivalent to increasing BNR by 10 dB between successive groups). The normalised performance is not significantly affected by the intrinsic noise, but the APoly and APLS methods appear to perform better than the other methods at the three levels tested. The trend for other peak to background ratios was similar.

Table 3. Median error ratios and quartile boundaries for background estimation using five different methods for a range of background to shot-noise ratios.

| Background to Shot Noise Ratio (BNR) | Method | Normalised Error Ratio | | |
|---|---|---|---|---|
| | | Lower Quartile | Median | Upper Quartile |
| 4.9 (13.8 dB) | ModPoly | 0.365 | 0.453 | 0.589 |
| | IModPoly | 0.253 | 0.454 | 0.792 |
| | APoly | 0.083 | 0.120 | 0.224 |
| | WPLS | 0.224 | 0.323 | 0.506 |
| | APLS | 0.049 | 0.066 | 0.092 |
| 15 (23.5 dB) | ModPoly | 0.221 | 0.390 | 0.654 |
| | IModPoly | 0.349 | 0.534 | 0.832 |
| | APoly | 0.086 | 0.202 | 0.360 |
| | WPLS | 0.310 | 0.310 | 0.684 |
| | APLS | 0.021 | 0.021 | 0.106 |
| 46.3 (33.3 dB) | ModPoly | 0.285 | 0.463 | 0.733 |
| | IModPoly | 0.367 | 0.601 | 0.871 |
| | APoly | 0.167 | 0.271 | 0.451 |
| | WPLS | 0.271 | 0.450 | 0.614 |
| | APLS | 0.019 | 0.040 | 0.099 |

### 3.1.5. Monotonic background signals

For simpler (monotonic polynomial) backgrounds for which the ModPoly, IModPoly and APoly methods are well adapted, APoly and APLS perform as well or better than the other methods tested (Fig 5), especially at the lower BNR tested. At the higher BNR the ModPoly method appears to outperform all of the other methods except APoly. It should be noted that the background test data matches the assumed form of the background used for the polynomial methods.[19]

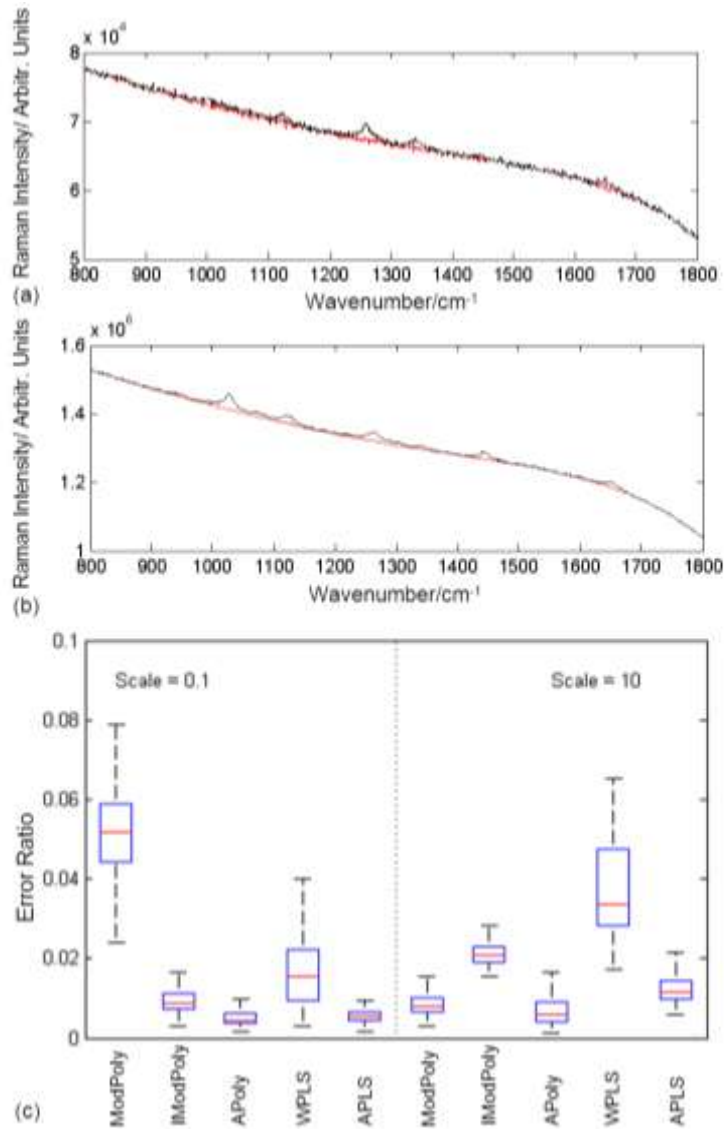

Figure 5. Samples from the polynomial background ensemble based on the test data of Zhao *et al*.[19]: (a) scale factor = 500; (b) scale factor = 10,000. (c) box plots for error ratios (at or close to optimum $\gamma$ for APLS and WPLS) for the randomised test data of Zhao *et al*. The results are sensitive to the scaling of the data, the scales here correspond to 500 and 10,000 times the raw spectrum (corresponding to an increase of 13 dB in BNR between the two cases). The background is a 5$^{th}$ order polynomial which closely matches the assumptions in the ModPoly, IModPoly and APoly methods.

*3.1.6. Summary*

In general the APoly and APLS methods described here appear to outperform the ModPoly, IModPoly and WPLS methods on spectra with complex backgrounds. For simpler backgrounds, at the lower BNR tested, APoly and APLS generally match or exceed the performance of the other methods. At the higher BNR tested, they exceed the performance of IModPoly and WPLS. The improvement observed in the ModPoly results at the higher BNR could be due to the close fit between the assumed background and the test background, but this has yet to be tested in detail.

As an aid to the interpretation of the box-plots presented above, a comprehensive set of plotted spectra and extracted backgrounds has been included in the on-line Supplementary Material. These plots (Figures S1 to S35) generally support the conclusions reached above based on the full ensembles.

**3.2. Experimental data**

The $O^2W^1$ background-correction result for the Raman spectra from a 1:8 mixture of dextran and fibrinogen using the proposed algorithm is shown in Fig 6. The adaptive weighting, combined with the penalised least squares estimation, automatically identifies likely background regions and pulls the background estimate below the data points.

The results of a comparative study using the 4 other background correction methods, (ModPoly, IModPoly, APoly and WPLS) are also shown in Fig 6. While lacking the quantitative support available with simulated data, for similar fitting parameter values, the results of the current algorithms (APoly and APLS) qualitatively compare favourably with those of the other methods tested, without, perhaps, some of the questionable features associated with the hard background and foreground classification used in those approaches.

For APLS, once an optimal roughness parameter has been selected, this single parameter can be kept fixed for consistent analysis of a series of related spectra.

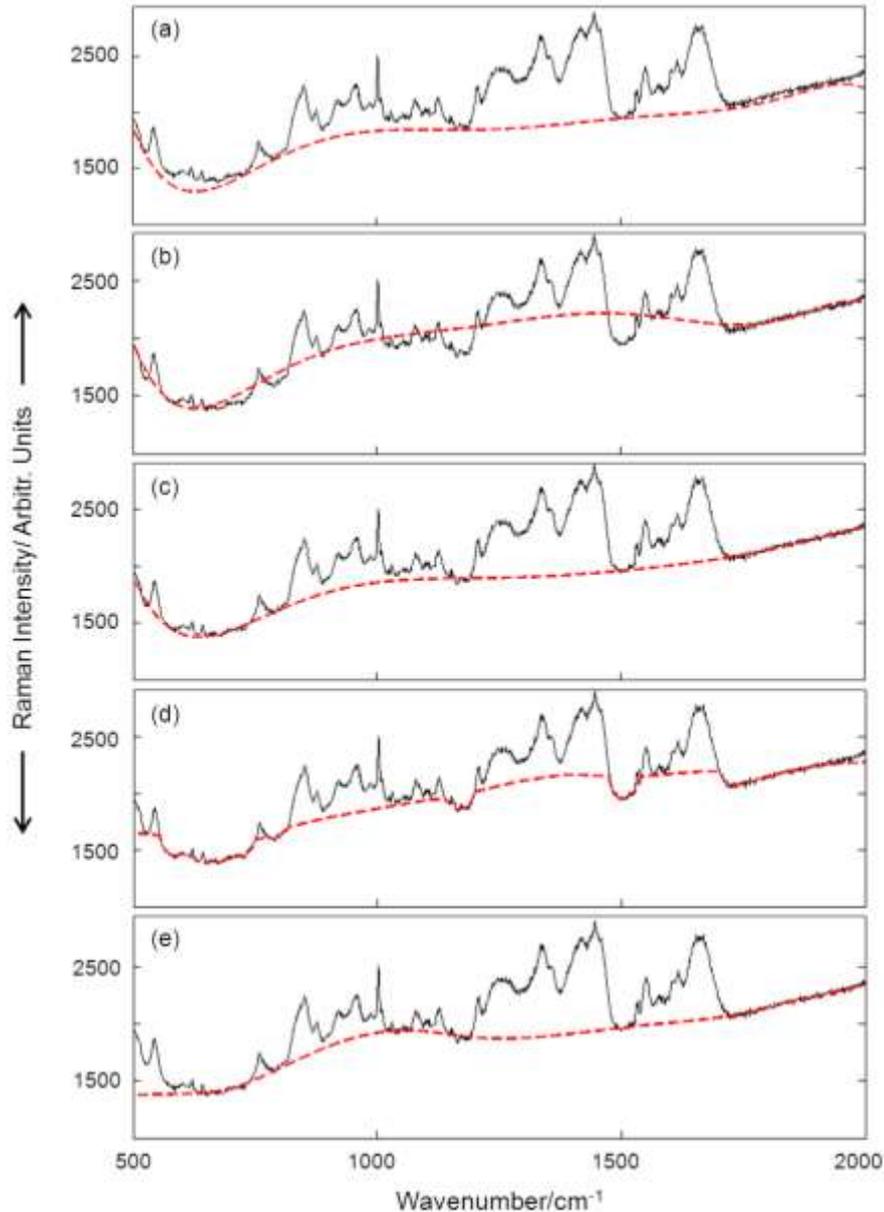

Figure 6. Experimental data fitted by (a) ModPoly (7$^{th}$ order, 250 iterations); (b) IModPoly (7$^{th}$ order, 250 iterations); (c) APoly (7$^{th}$ order, 250 iterations); (d) WPLS ($\gamma = 75,000$); (e) APLS ($\gamma = 750,000$). The extracted background is the dashed line (red on-line).

## 4. Conclusions

Methods are reported for improving fluorescence background removal using probability based adaptive-weights. A single, relatively simple criterion is used to adaptively determine the weights used in penalised least squares and polynomial estimations. The weights are based on the probability that a point is part of the background given the Poisson statistics of the signal. Compared to related methods that use a simple binary classification of peaks and background, this continuously variable weighting helps to improve the background estimation. The results from simulated spectra with artificial background demonstrate the efficiency and accuracy of the proposed algorithm over the range of signal parameters tested. Over this range it generally provides a significant improvement on the polynomial baseline subtraction methods for structured background spectra. The adaptive weight, penalized least squares approach also generally improves on the method of Zhang *et al*. [22] while avoiding questionable features associated with rapid changes in slope of the fitting curve (see Fig. 6(d) and Supplementary Material figures S4, S9, S14, S19, S24, S29 and S34 for examples). With application of the present algorithm, consistent Raman spectra with significantly improved background can reasonably be expected to be obtained without additional pre-processing.

The main remaining problem, common to PLS methods in general, is to find a reliable method for automatically setting the smoothing factor, $\gamma$. Fortunately, as shown in Figure 2(b) (and other results not shown here), the range of values for $\gamma$, over which good performance can be expected for spectra of a given type, appears to be reasonably large so that interactive fitting of typical spectra remains a viable alternative in practice.

**Acknowledgment**

The authors thank Dr. Michelle Dunn for advice in MATLAB analysis.

# References

[1] J. Chan, S. Fore, S. Wachsman-Hogiu, T. Huser, *Laser & Photon. Rev.* **2008**, *2*, 325.
[2] W. E. Huang, R. I. Griffiths, I. P. Thompson, M. J. Bailey, A. S. Whiteley, *Anal. Chem.* **2004**, *76*, 4452.
[3] M. A. Strehle, P. Rosch, R. Petry, A. Hauck, R. Thull, W. Kiefer, J. Popp, *Phys. Chem. Chem. Phys.* **2004**, *6*, 5232.
[4] V. Mazet, C. Carteret, D. Brie, J. Idier, B. Humbert, *Chemometr. Intell. Lab.* **2005**, *7*, 121.
[5] G. Schulze, A. Jirasek, M. M. L. Yu, A. Lim, R. F. B. Turner, M. W. Blades, *Appl. Spectrosc.* **2005**, *59*, 545.
[6] A. P. Shreve, N. J. Cherepy, R. A. Mathies, *Appl. Spectrosc.* **1992,** *46*, 707.
[7] J. J. Baraga, M. S. Feld, R. P. Rava, *Appl. Spectrosc.* **1992**, *46*, 187.
[8] D. L. A. de Faria, M. A. de Souza, *J. Raman Spectrosc.* **1999**, *30*, 169.
[9] P. Matousek, M. Towrie, A. W. Parker, *Appl. Spectrosc.* **2005**, *59*, 848.
[10] F. Knorr, Z. J. Smith, S. Wachsmann-Hogiu, *Opt. Express.* **2010**, *18*, 20049.
[11] P. Matousek, M. Towrie, C. Ma, W. M. Kwok, D. Phillips, W. T. Toner, A. W. Parker, *J. Raman Spectrosc.* **2001,** *32*, 983.
[12] A. M. Macdonald, P. Wyeth, *J. Raman Spectrosc.* **2006**, *37*, 830.
[13] A. P. Esposito, C. E. Talley, T. Huser, C. W. Hollars, C. M. Schaldach, S. M. Lane, *Appl. Spectrosc.* **2003,** *57*, 868.
[14] F. Bonnier, S. M. Ali, P. Knief, H. Lambkin, K. Flynn, V. McDonagh, C. Healy, T. C. Lee, F. M. Lyng, H. J. Byrne, *Vib. Spectrosc.* **2012,** *61*, 124.
[15] A. O'Grady, A. C. Dennis, D. Denvir, J. J. McGarvey, S. E. J. Bell, *Anal. Chem.* **2001**, *73*, 2058.
[16] D. M. Zhang, D. Ben-Amotz, *Appl. Spectrosc.* **2000**, *54*, 1379.
[17] P. A. Mosierboss, S. H. Lieberman, R. Newbery, *Appl. Spectrosc.* **1995**, *49*, 630.
[18] A. Mahadevan-Jansen, M. F. Mitchell, N. Ramanujam, A. Malpica, S. Thomsen, U. Utzinger, R. Richards-Kortum, *Photochem. Photobiol.* **1998**, *68*, 123.
[19] J. Zhao, H. Lui, D. I. McLean, H. Zeng, *Appl. Spectrosc.* **2007**, *61*, 1225.
[20] J. F. Brennan, Y. Wang, R. R. Dasari, M. S. Feld, *Appl. Spectrosc.* **1997**, *51*, 201.
[21] T. T. Cai, D. M. Zhang, D. Ben-Amotz, *Appl. Spectrosc.* **2001**, *55*, 1124.
[22] Z. M. Zhang, S. Chen, Y. Z. Liang, Z. X. Liu, Q. M. Zhang, L. X. Ding, F. Ye, H. Zhou, *J. Raman Spectrosc.* **2010**, *41*, 659.
[23] C. A. Lieber, A. Mahadevan-Jansen, *Appl. Spectrosc*, **2003**, *57*, 1363.
[24] http://code.google.com/p/baselinewavelet/, accessed 5 Dec 2012. (See also [22]).
[25] Note that in equation (5) of Zhao et al [19] which describes their test background, there appears to be a typographical error in the zero order coefficient, which was corrected (to 382.2) in calculating the ensembles used here.

Supplementary Material

Figures S1 to S35 contain plots of the test data, the background fits and the recovered spectra obtained under a range of ensemble parameter settings for each of the five methods described in the text of the main paper. The three columns of each plot correspond to plots selected to be at or close to the upper bound of the lower quartile (first column), at the median (second column) and at or close to the lower bound of the upper quartile (third column) of the error ratios calculated from Equation (8). The corresponding error ratios can be found in Tables 2 and 3 in the main paper. The top row in each figure shows the simulated test data (Raman spectrum plus background) in black and the simulated background in red. The middle row shows the simulated background (black) and the extracted mean background using the method indicated in the caption in red. The lower row shows the simulated Raman signal (black) and the spectrum recovered by background subtraction using the extracted background (red).

The variable parameters for the ensembles used are:

(1)   The average number of peaks in the scan range (Np).

(2).   The range of spectral peak heights over which the peaks are randomly uniformly distributed, $[P_1, P_2]$.

(3).   The range of amplitudes over which the 3 sinusoidal components of the simulated background are randomly uniformly distributed, $[A_1, A_2]$.

(4).   The overall scaling of the data which sets the level of mean background to shot noise ratio, (Sf, BNR)

(5).   The method used to extract the background, (METHOD). Where appropriate the variable parameters for the fitting methods have been optimised independently for each member of an ensemble

The appropriate values of these parameters are given in the captions to each of the figures.

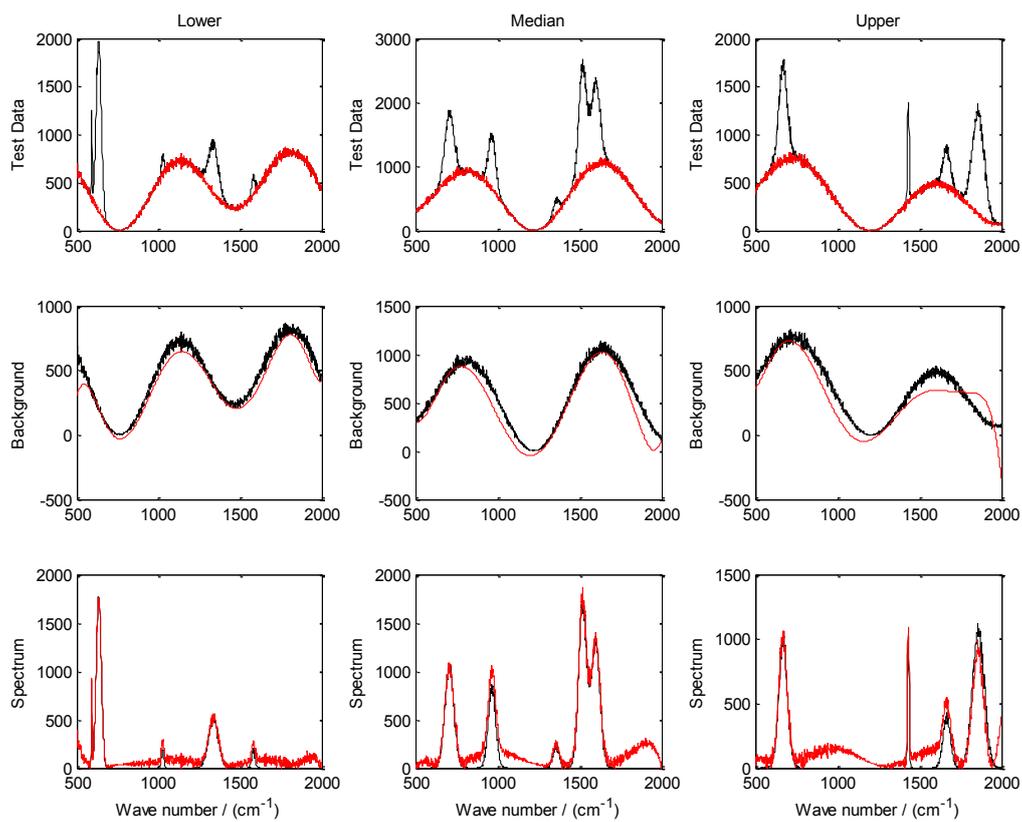

Figure S1.　　Np = 5; P$_1$ = 0; P$_2$ = 2000; A$_1$ = 0; A$_2$ = 400; Sf = 1 (BNR = 15); METHOD = ModPoly.

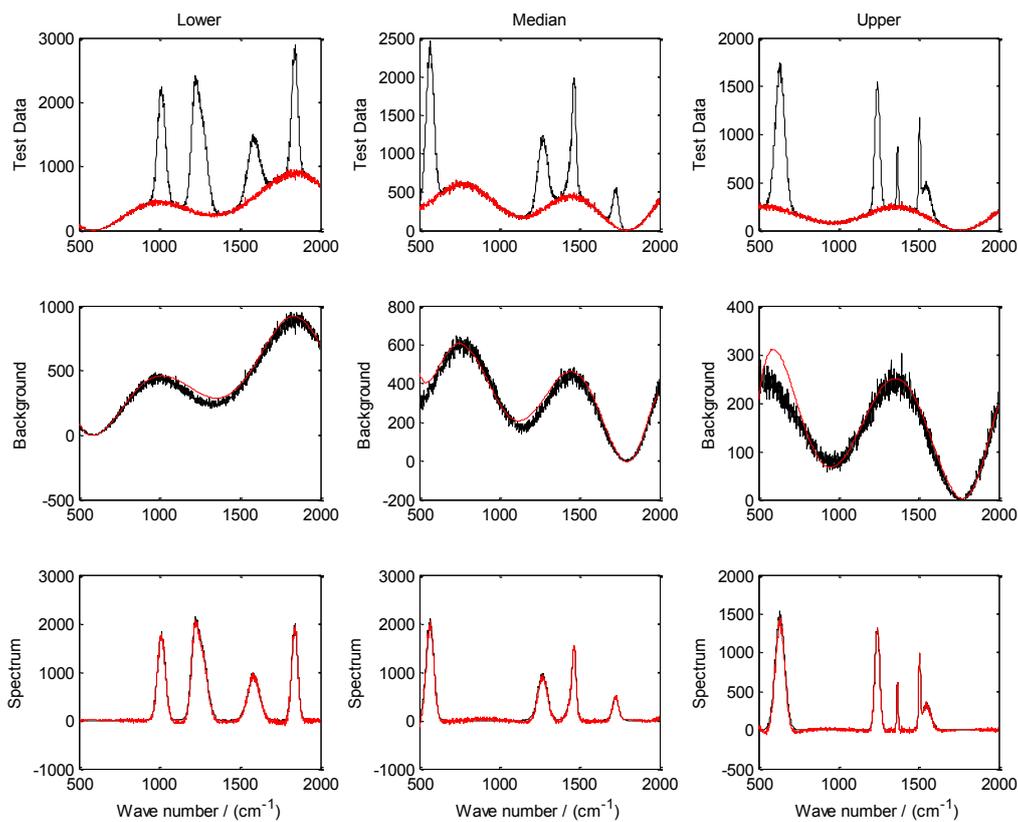

Figure S2.     $Np = 5$; $P_1 = 0$; $P_2 = 2000$; $A_1 = 0$; $A_2 = 400$; $Sf = 1$ (BNR = 15); METHOD = IModPoly.

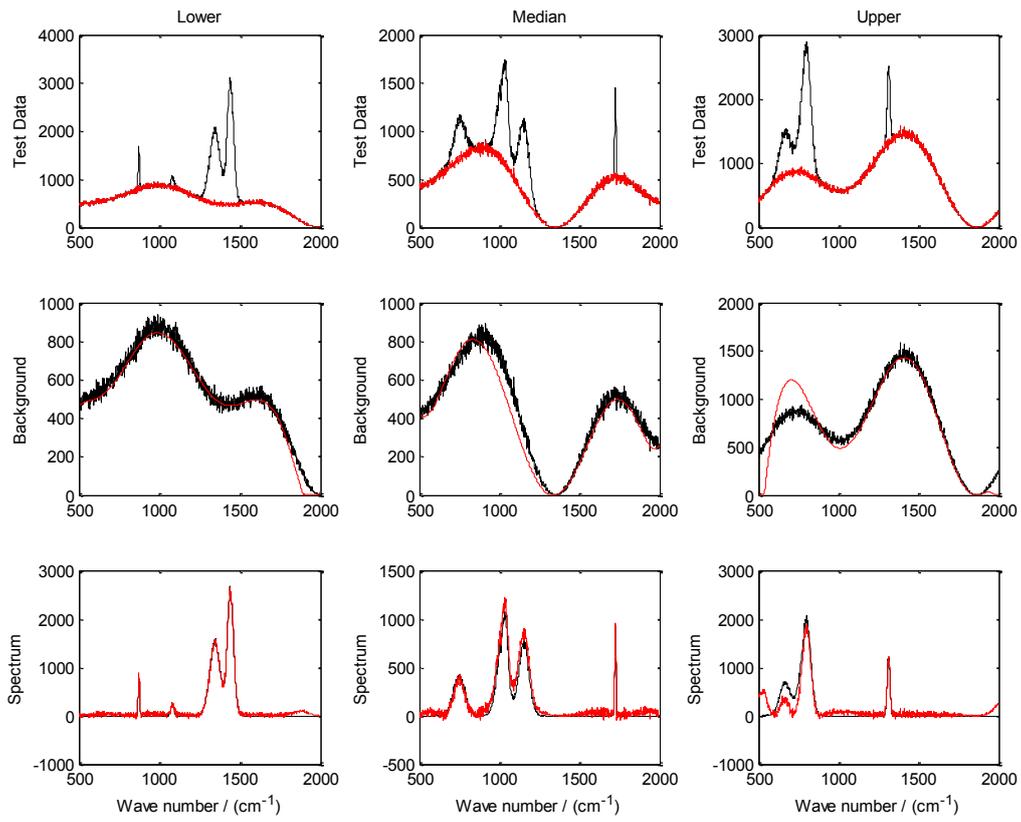

Figure S3.  Np = 5; $P_1$ = 0; $P_2$ = 2000; $A_1$ = 0; $A_2$ = 400; Sf = 1 (BNR = 15); METHOD = APoly.

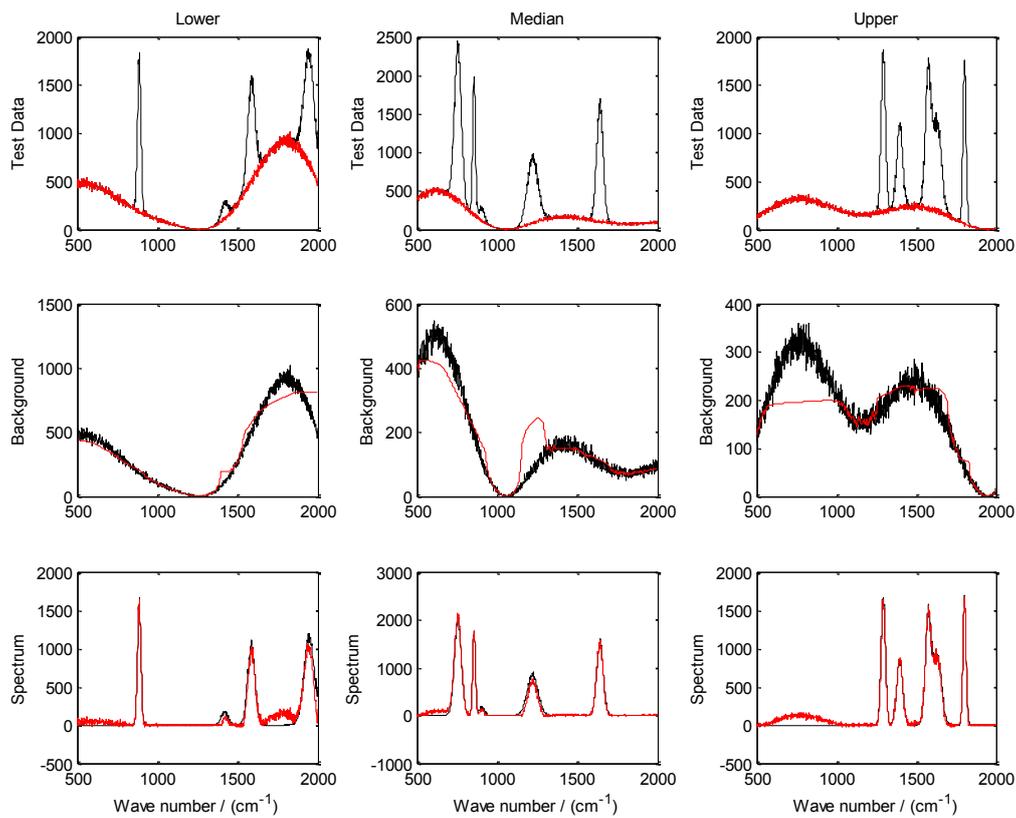

Figure S4. Np = 5; $P_1$ = 0; $P_2$ = 2000; $A_1$ = 0; $A_2$ = 400; Sf = 1 (BNR = 15); METHOD = WPLS.

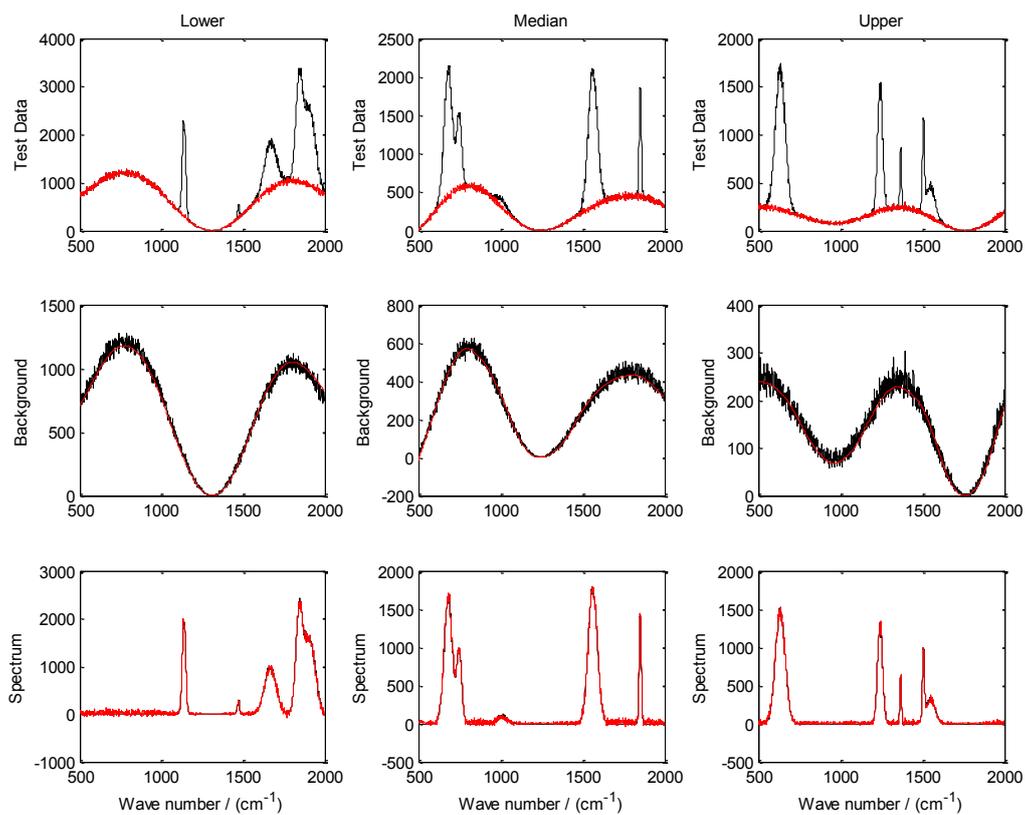

Figure S5. Np = 5; $P_1$ = 0; $P_2$ = 2000; $A_1$ = 0; $A_2$ = 400; Sf = 1 (BNR = 15); METHOD = APLS.

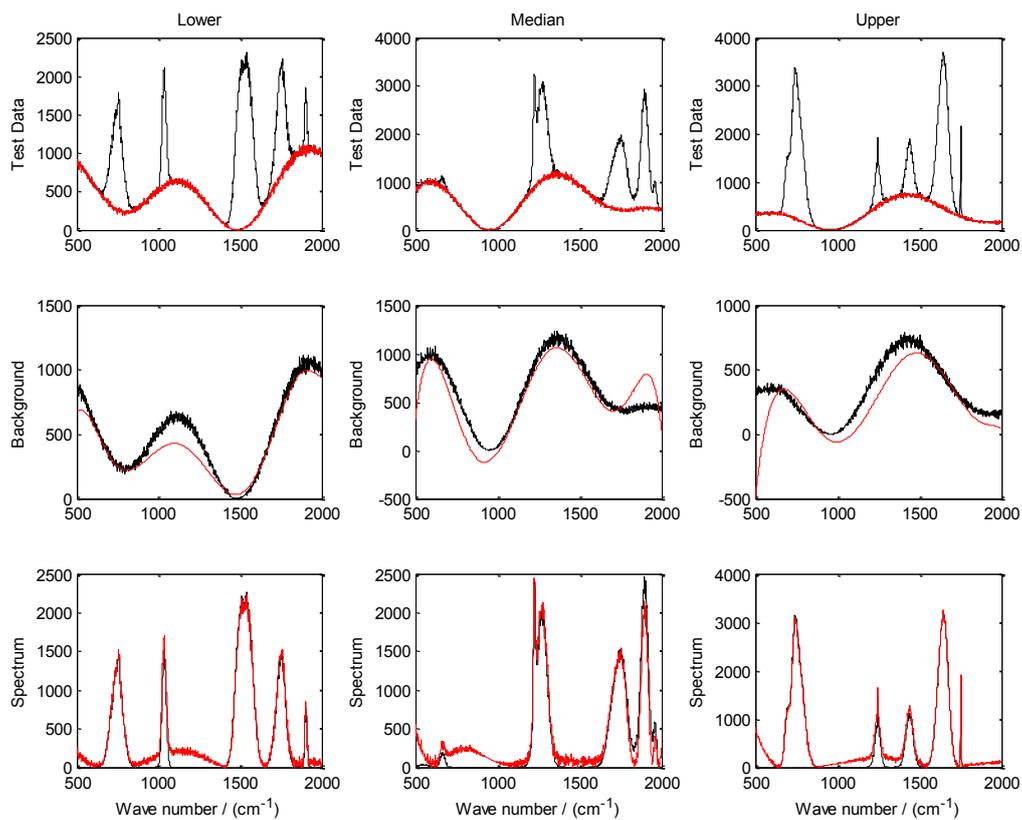

Figure S6.   Np = 10; $P_1$ = 0; $P_2$ = 2000; $A_1$ = 0; $A_2$ = 400; Sf = 1 (BNR = 15); METHOD = ModPoly.

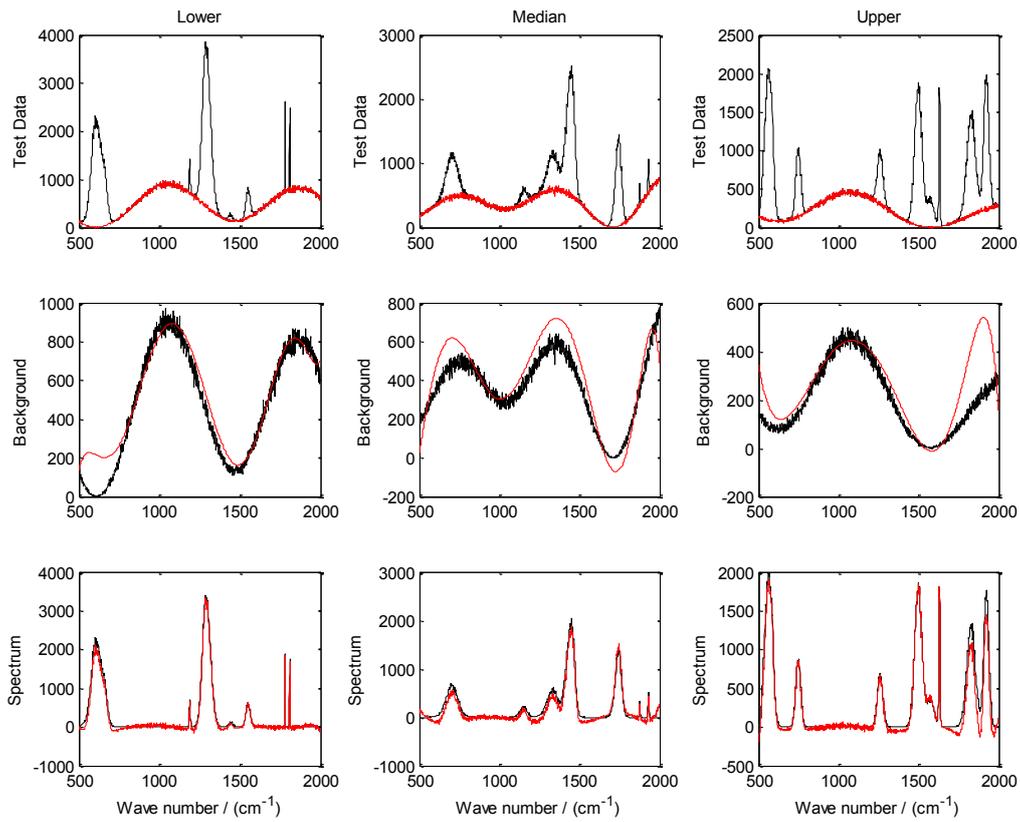

Figure S7. $Np = 10$; $P_1 = 0$; $P_2 = 2000$; $A_1 = 0$; $A_2 = 400$; $Sf = 1$ (BNR = 15); METHOD = IModPoly.

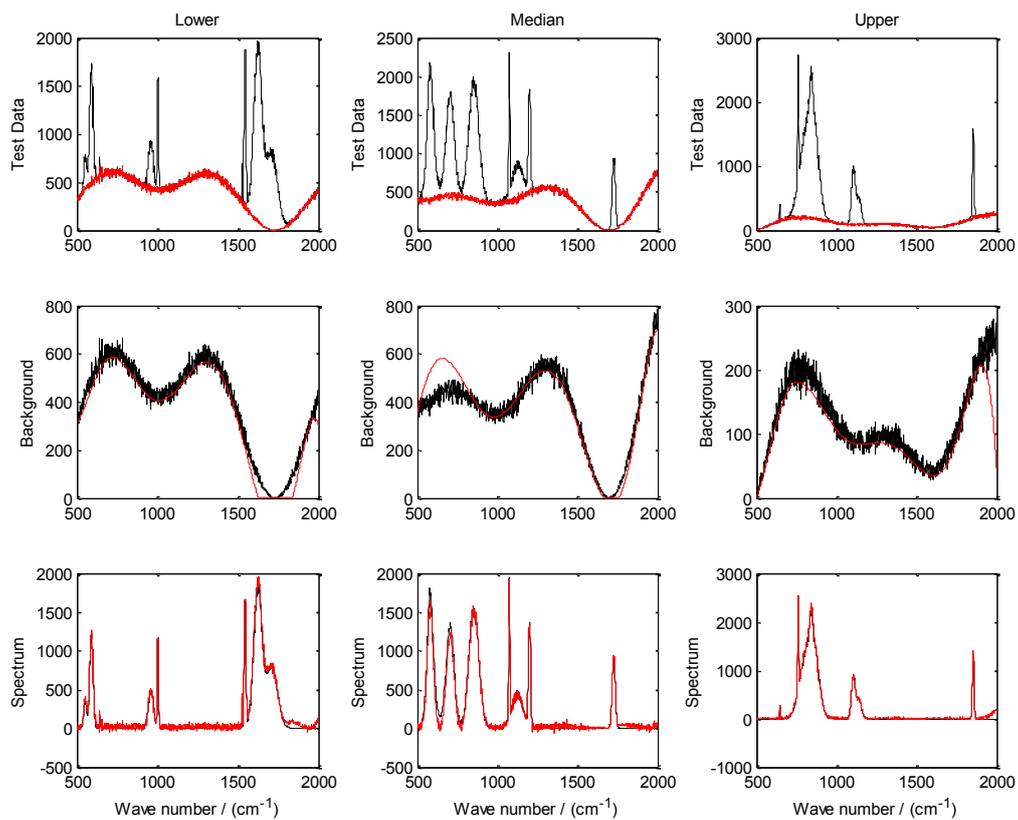

Figure S8. Np = 10; $P_1$ = 0; $P_2$ = 2000; $A_1$ = 0; $A_2$ = 400; Sf = 1 (BNR = 15); METHOD = APoly.

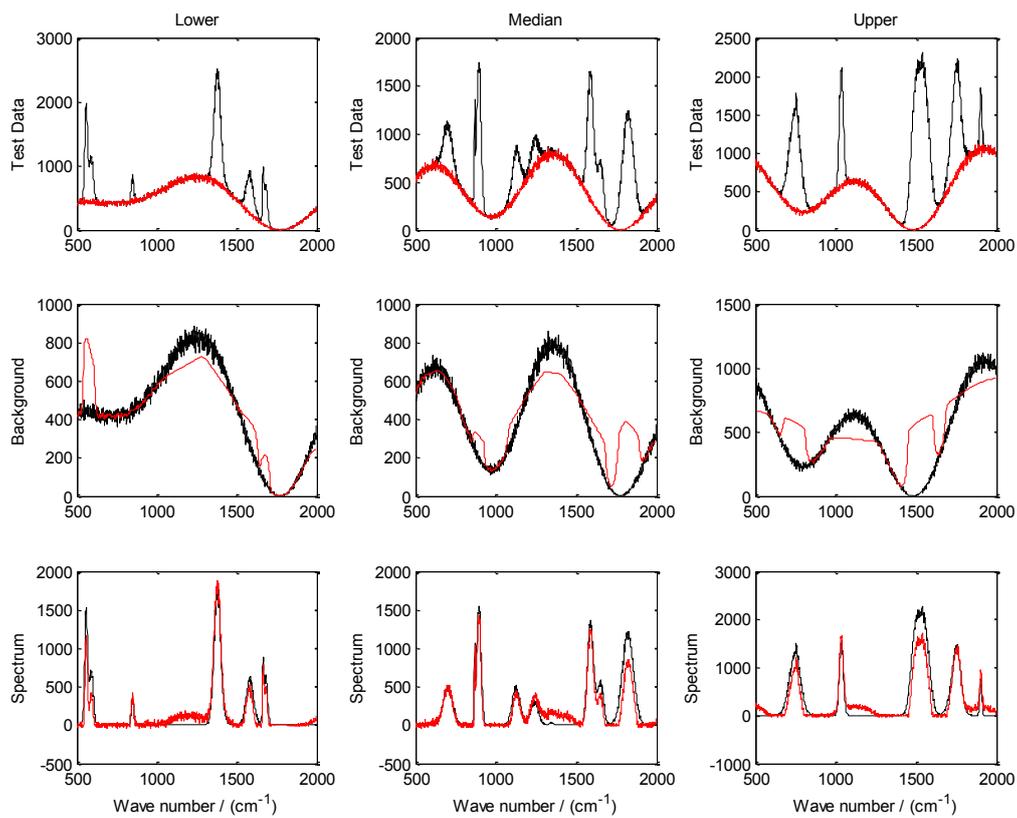

Figure S9. Np = 10; $P_1$ = 0; $P_2$ = 2000; $A_1$ = 0; $A_2$ = 400; Sf = 1 (BNR = 15); METHOD = WPLS.

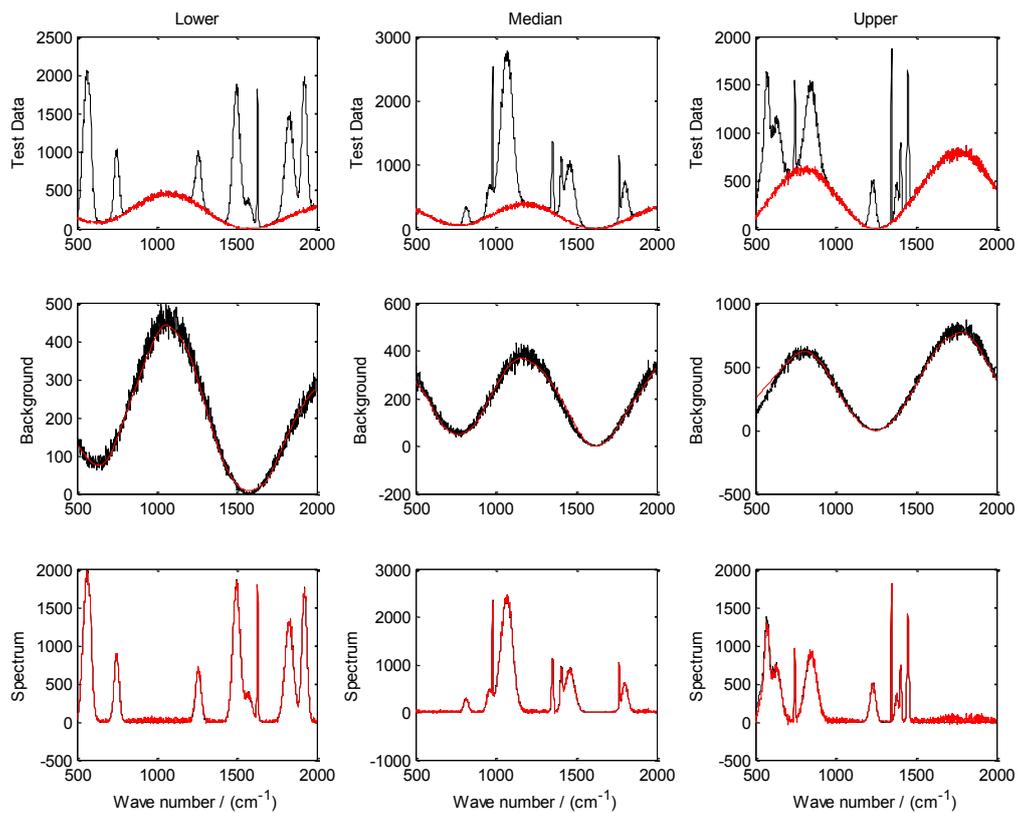

Figure S10. Np = 10; $P_1$ = 0; $P_2$ = 2000; $A_1$ = 0; $A_2$ = 400; Sf = 1 (BNR = 15); METHOD = APLS.

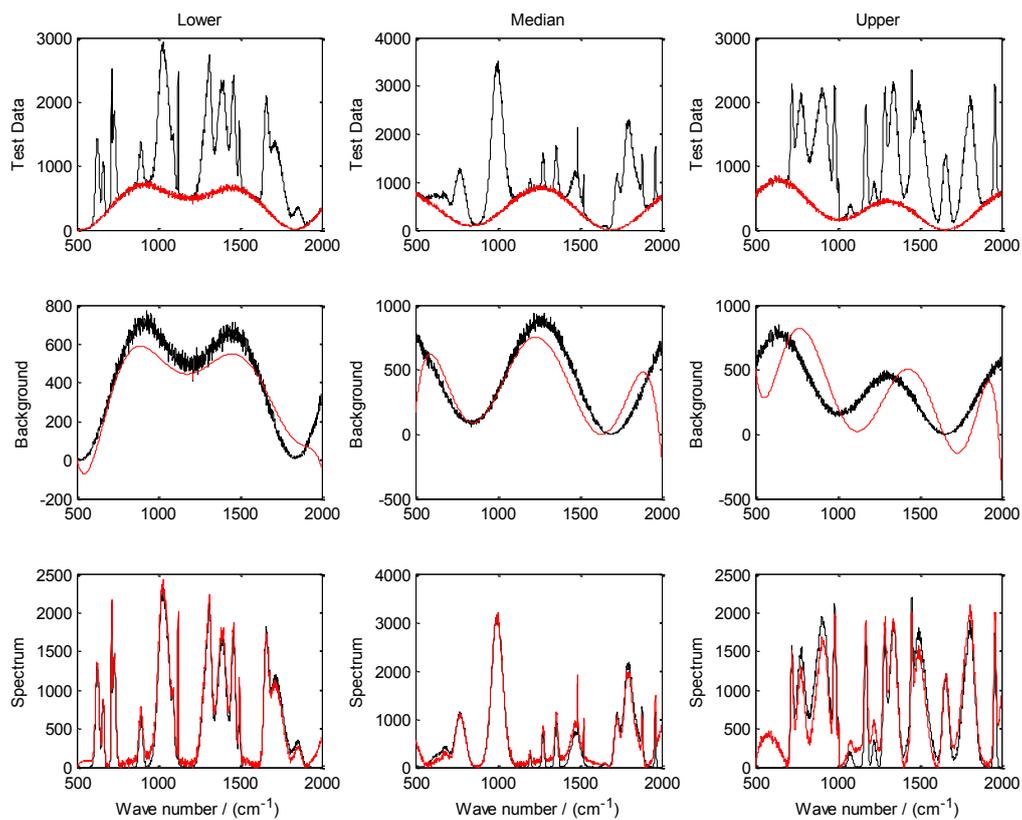

Figure S11.  Np = 20; $P_1$ = 0; $P_2$ = 2000; $A_1$ = 0; $A_2$ = 400; Sf = 1 (BNR = 15); METHOD = ModPoly.

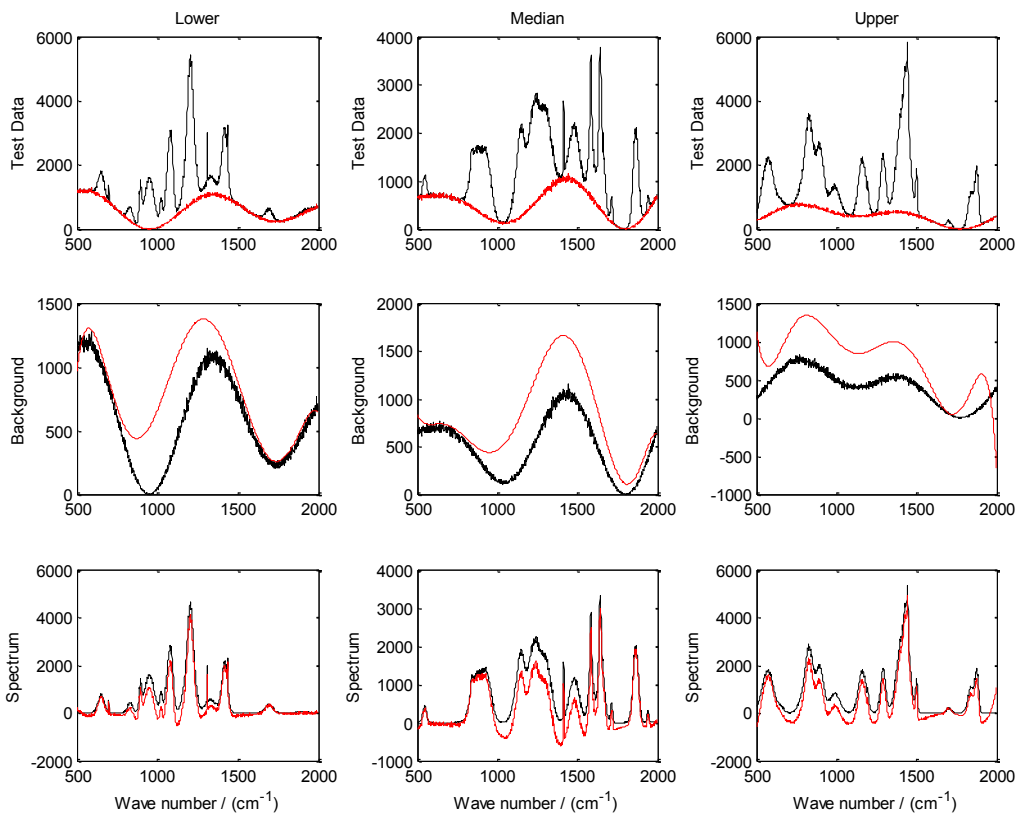

Figure S12. Np = 20; $P_1$ = 0; $P_2$ = 2000; $A_1$ = 0; $A_2$ = 400; Sf = 1 (BNR = 15); METHOD = IModPoly.

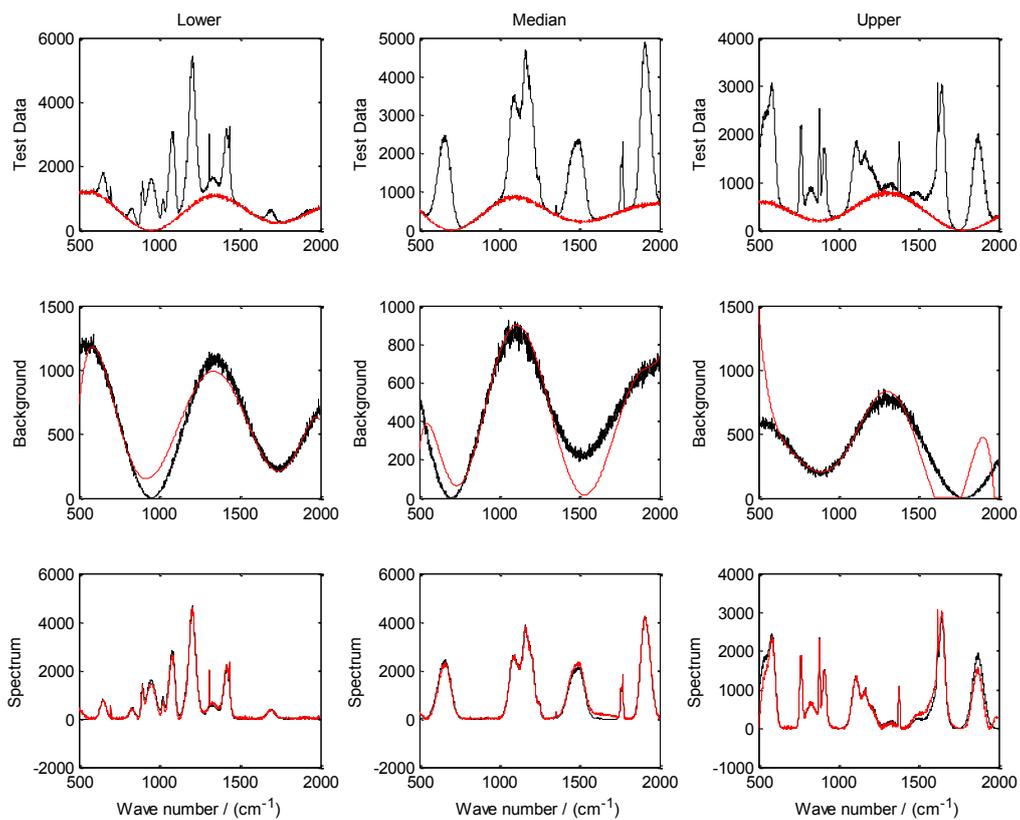

Figure S13. Np = 20; $P_1$ = 0; $P_2$ = 2000; $A_1$ = 0; $A_2$ = 400; Sf = 1 (BNR = 15); METHOD = APoly.

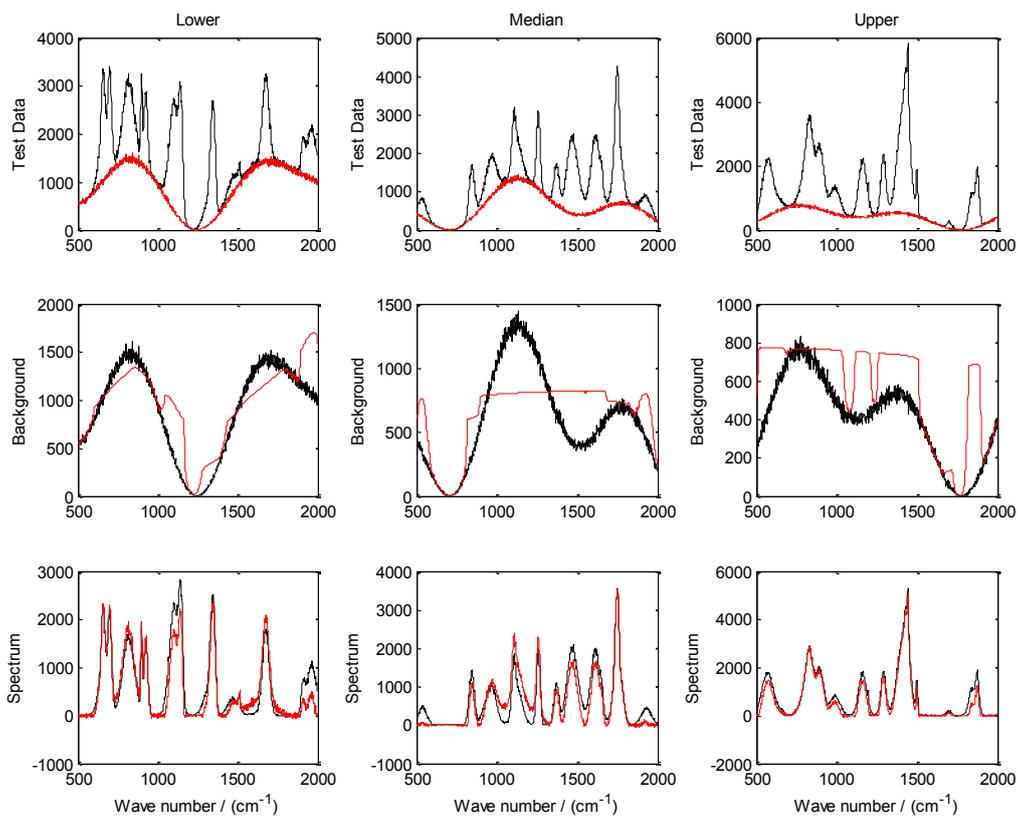

Figure S14.  Np = 20; $P_1$ = 0; $P_2$ = 2000; $A_1$ = 0; $A_2$ = 400; Sf = 1 (BNR = 15); METHOD = WPLS.

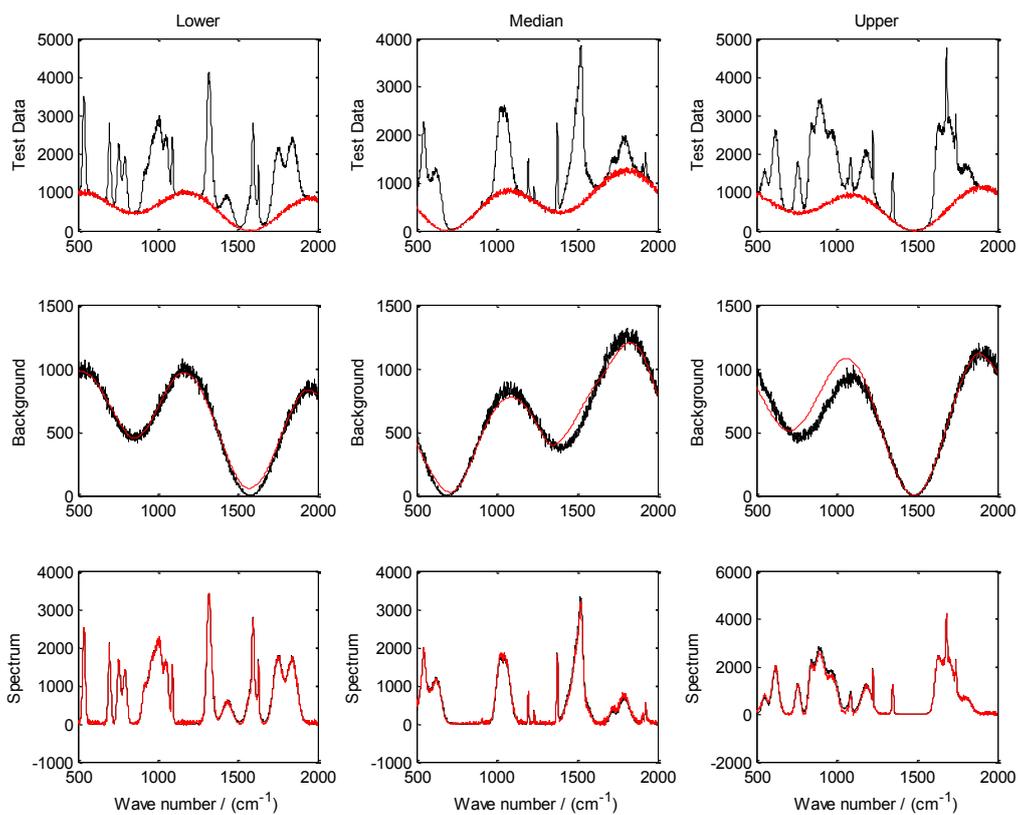

Figure S15. Np = 20; $P_1$ = 0; $P_2$ = 2000; $A_1$ = 0; $A_2$ = 400; Sf = 1 (BNR = 15); METHOD = APLS.

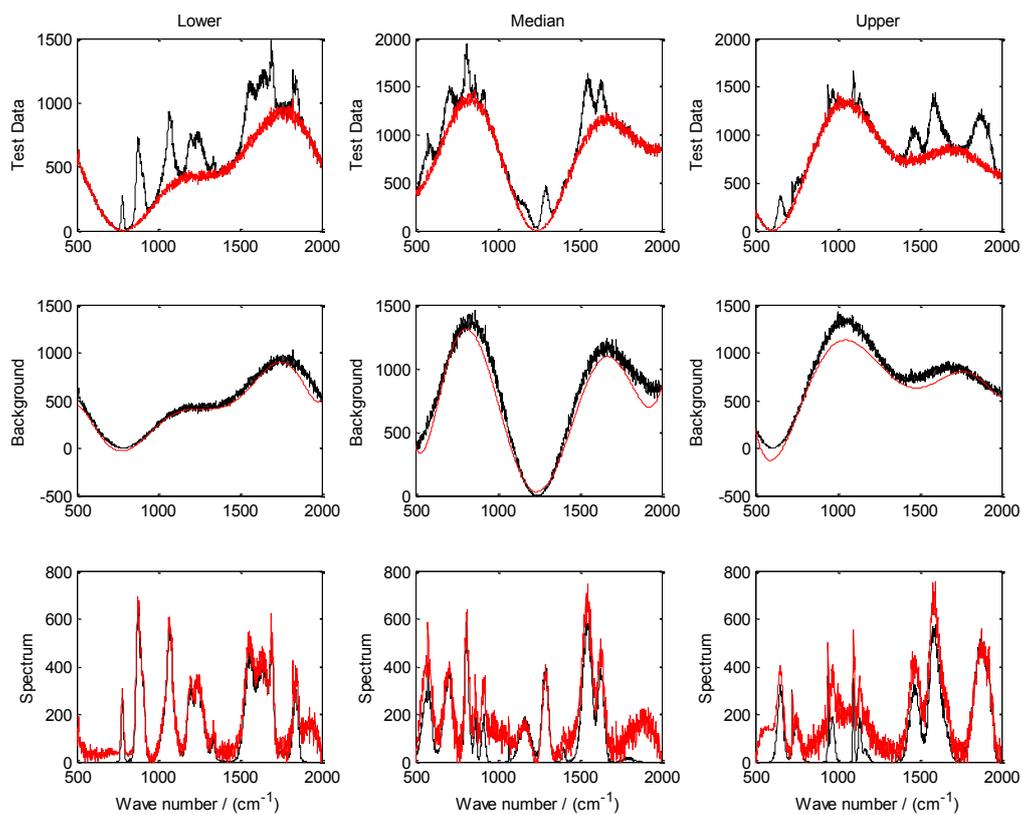

Figure S16. Np = 20; $P_1$ = 0; $P_2$ = 400; $A_1$ = 0; $A_2$ = 400; Sf = 1; METHOD = ModPoly.

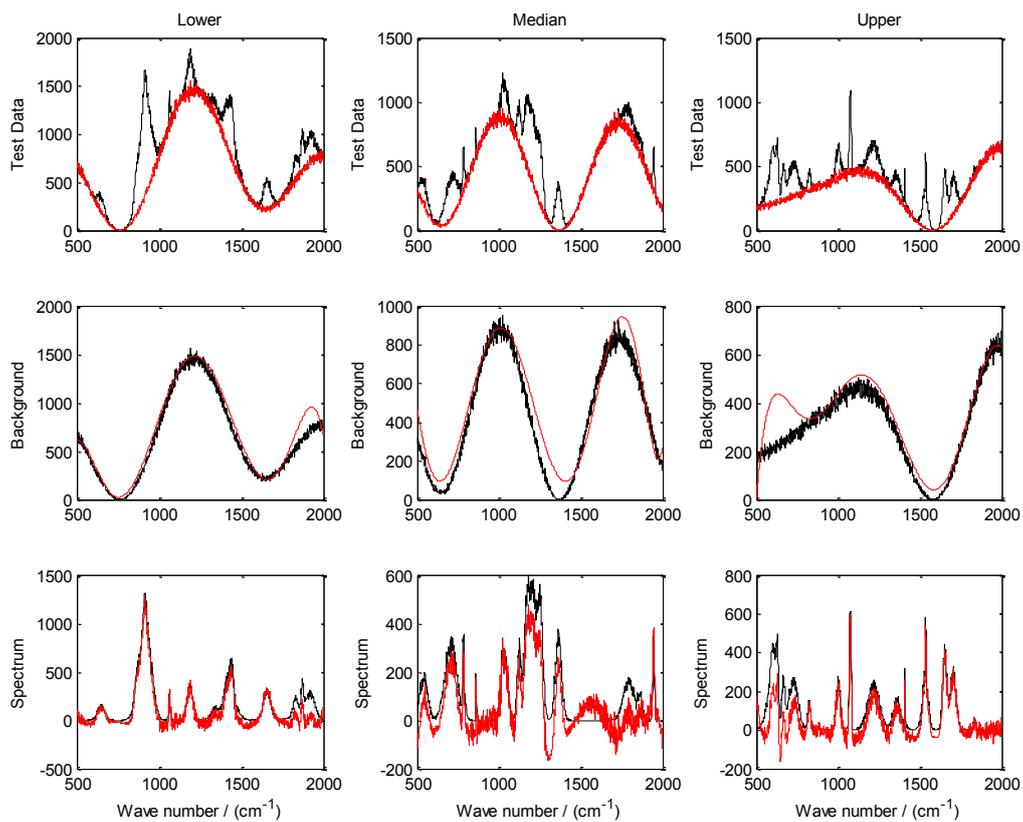

Figure S17. Np = 20; $P_1$ = 0; $P_2$ = 400; $A_1$ = 0; $A_2$ = 400; Sf = 1; METHOD = IModPoly.

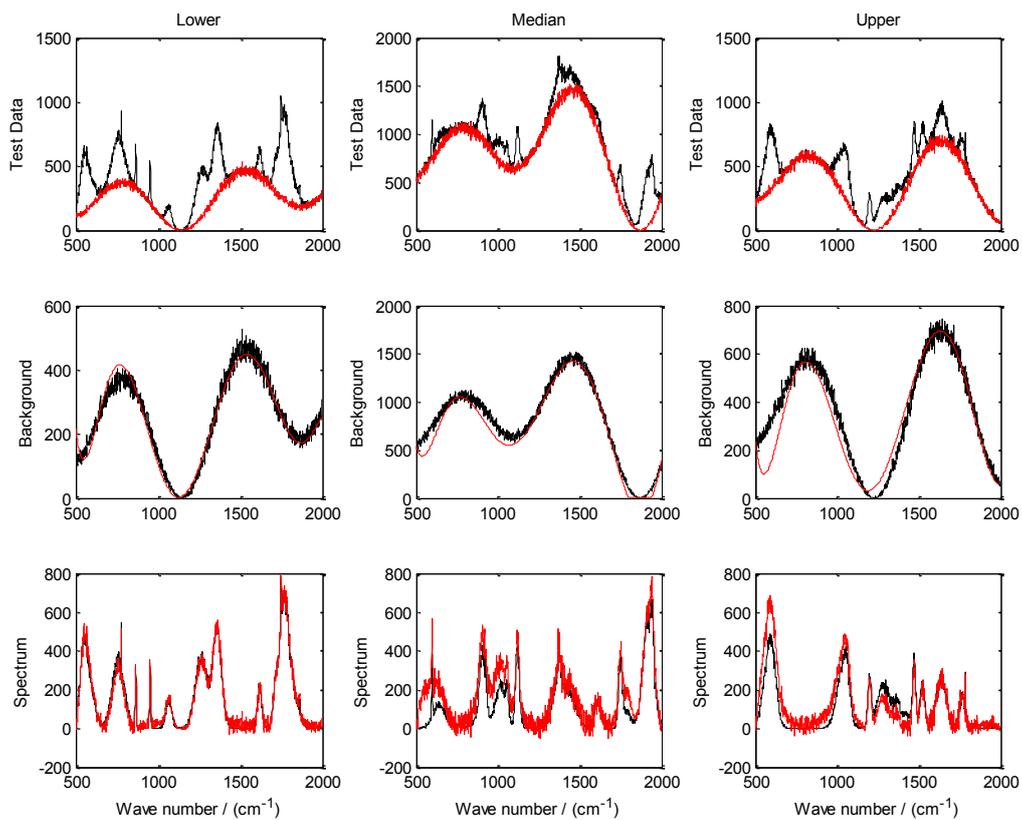

Figure S18. Np = 20; $P_1$ = 0; $P_2$ = 400; $A_1$ = 0; $A_2$ = 400; Sf = 1; METHOD = APoly.

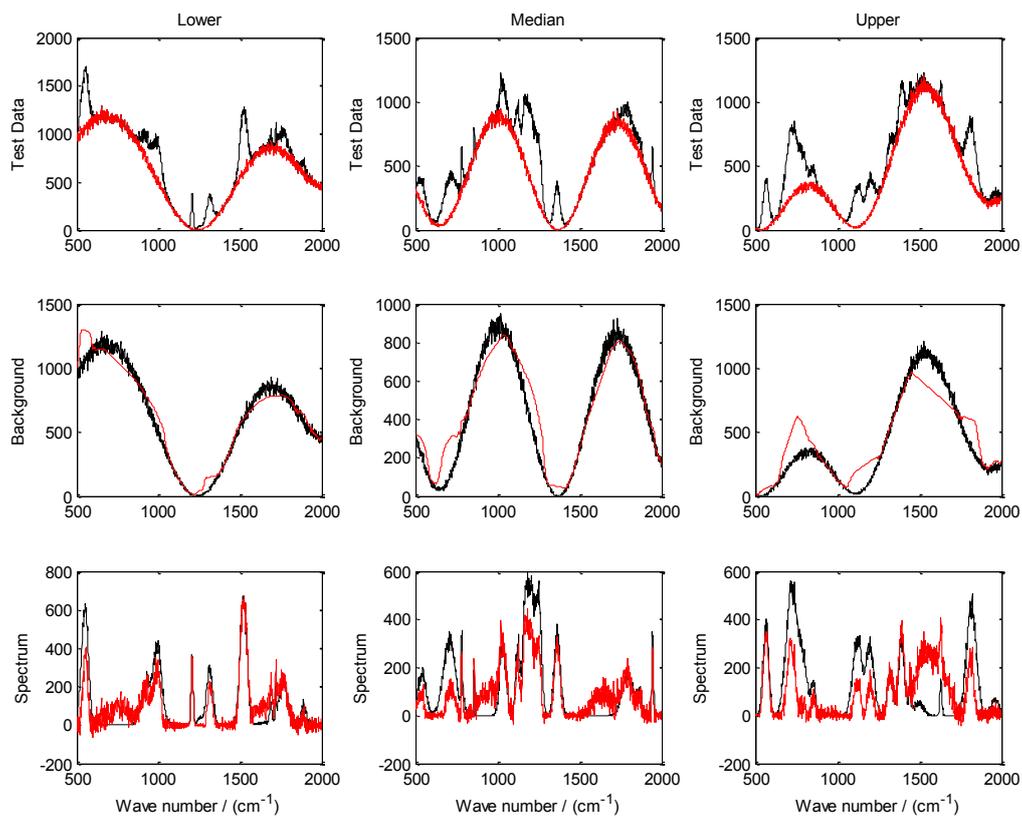

Figure S19.  Np = 20; $P_1$ = 0; $P_2$ = 400; $A_1$ = 0; $A_2$ = 400; Sf = 1; METHOD = WPLS.

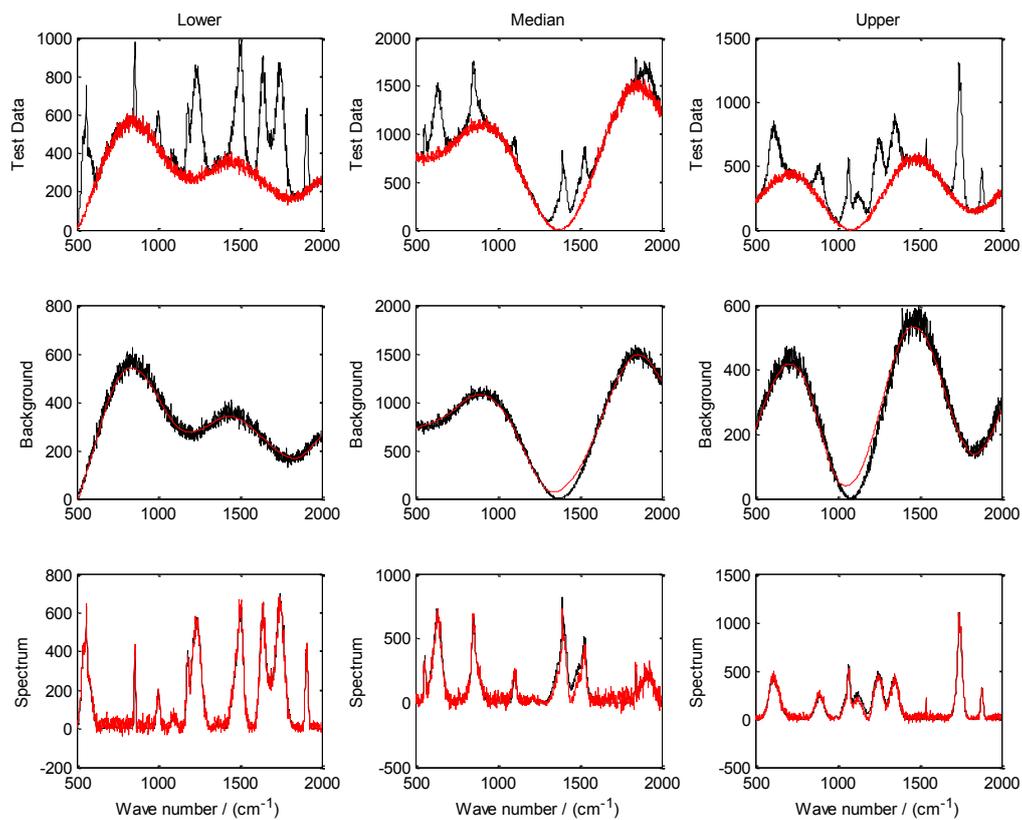

Figure S20.  Np = 20; $P_1$ = 0; $P_2$ = 400; $A_1$ = 0; $A_2$ = 400; Sf = 1; METHOD = APLS.

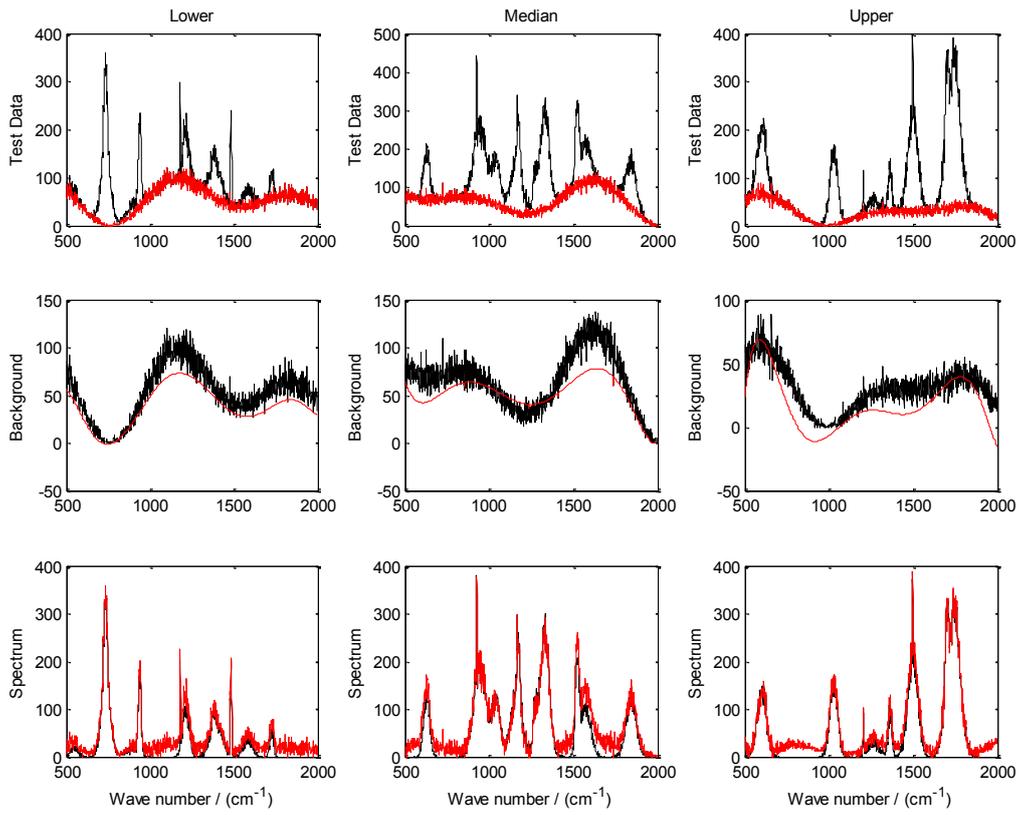

Figure S21. Np = 12; $P_1$ = 0; $P_2$ = 2000; $A_1$ = 0; $A_2$ = 400; Sf = 0.1 (BNR = 4.9); METHOD = ModPoly.

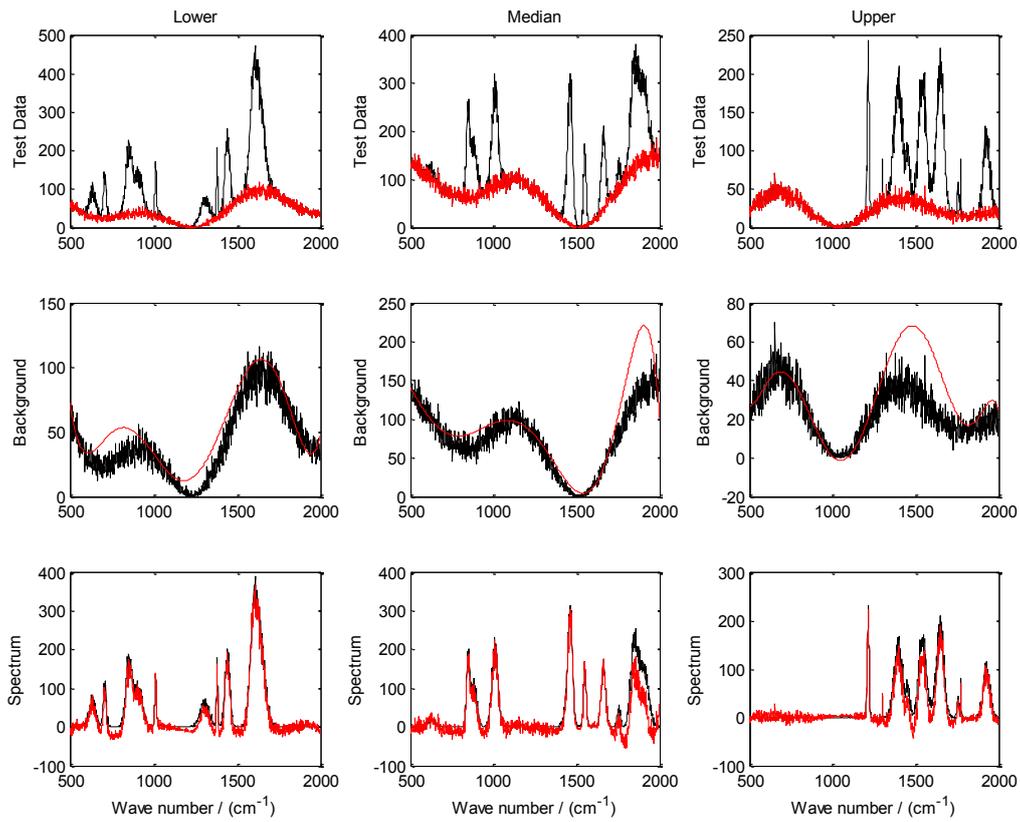

Figure S22.   Np = 12; $P_1$ = 0; $P_2$ = 2000; $A_1$ = 0; $A_2$ = 400; Sf = 0.1 (BNR = 4.9); METHOD = IModPoly.

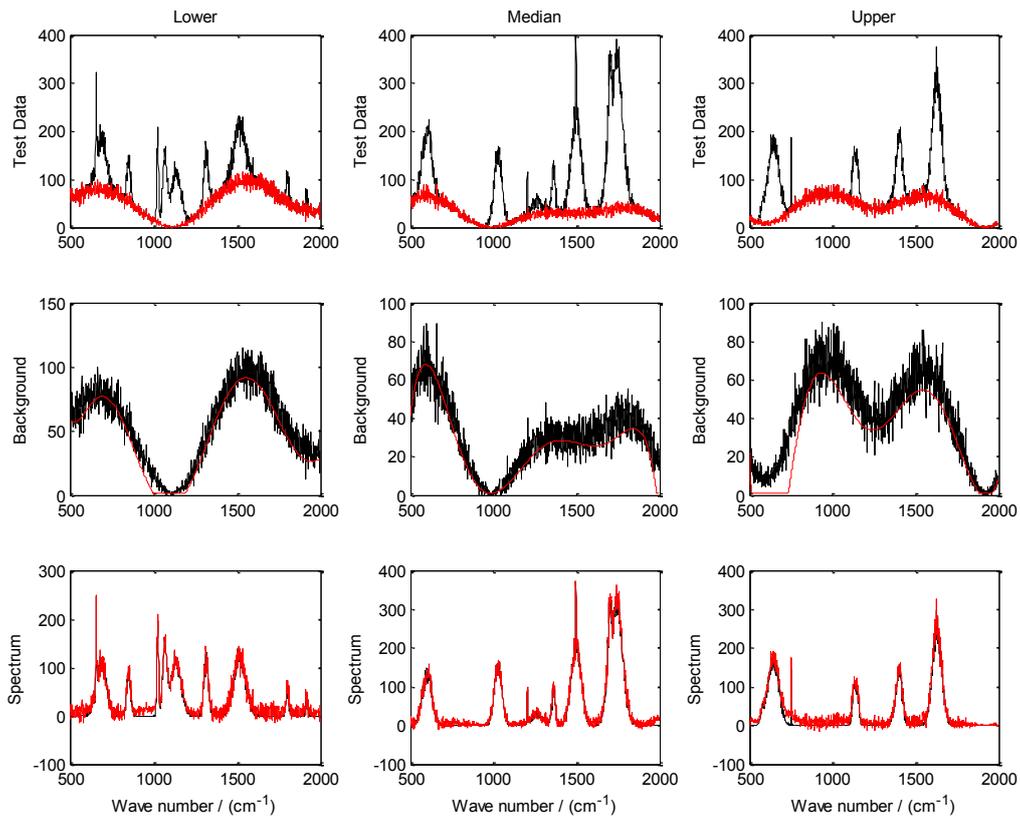

Figure S23. $N_p = 12$; $P_1 = 0$; $P_2 = 2000$; $A_1 = 0$; $A_2 = 400$; $S_f = 0.1$ (BNR = 4.9); METHOD = APoly

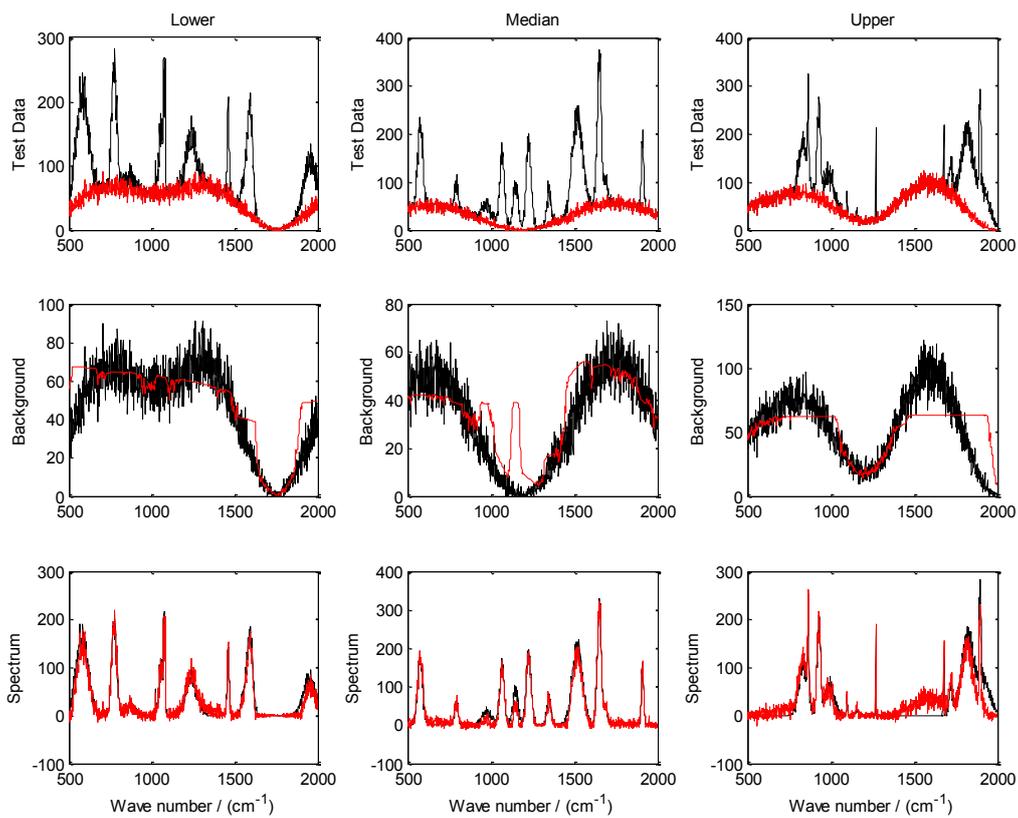

Figure S23.  Np = 12; $P_1$ = 0; $P_2$ = 2000; $A_1$ = 0; $A_2$ = 400; Sf = 0.1 (BNR = 4.9); METHOD = WPLS

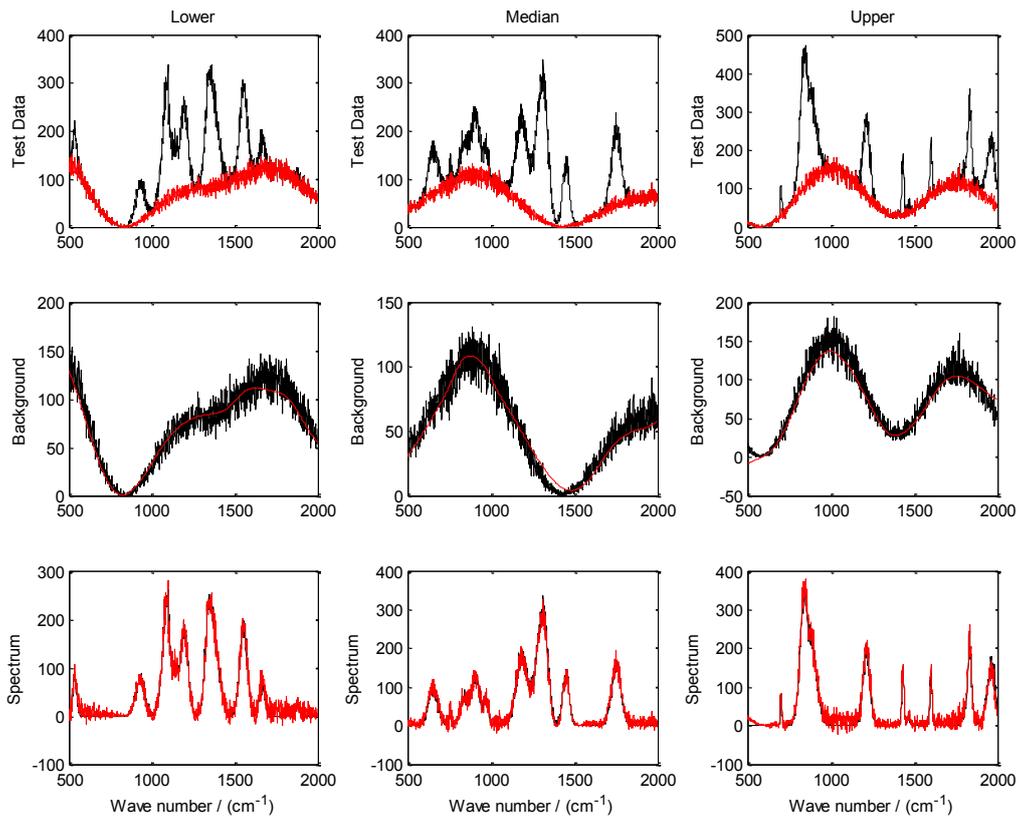

Figure S25.  Np = 12; $P_1$ = 0; $P_2$ = 2000; $A_1$ = 0; $A_2$ = 400; Sf = 0.1 (BNR = 4.9); METHOD = APLS

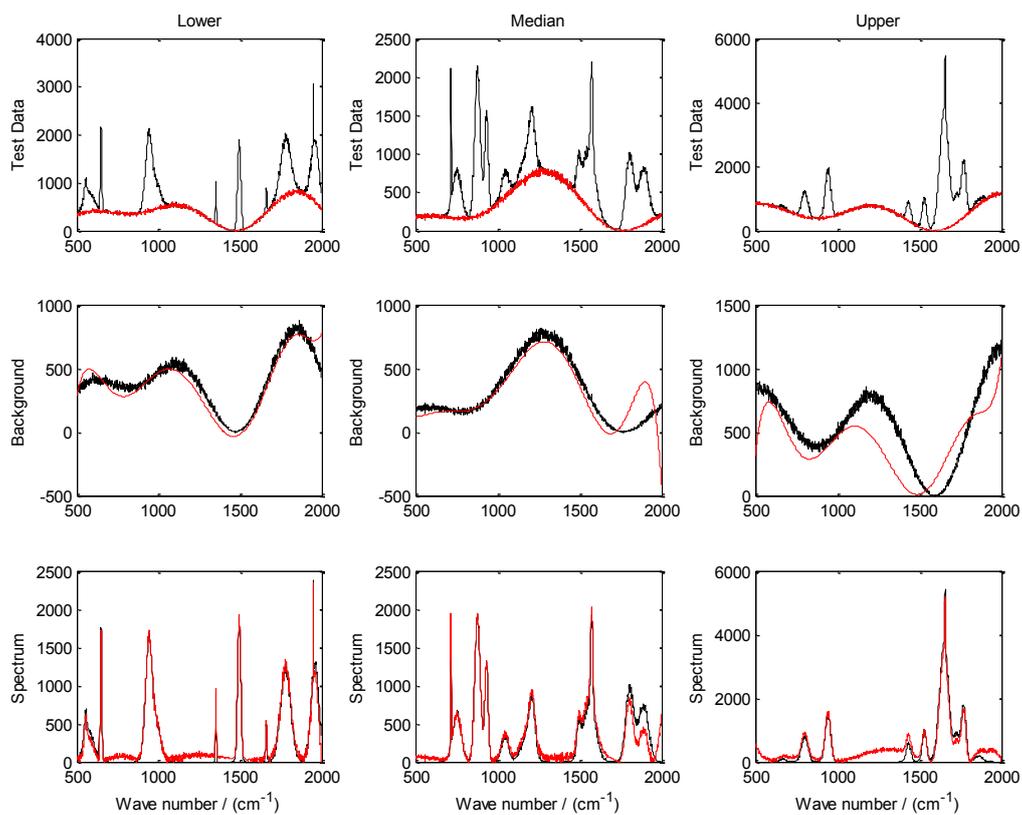

Figure S26. Np = 12; $P_1$ = 0; $P_2$ = 2000; $A_1$ = 0; $A_2$ = 400; Sf = 1 (BNR = 15); METHOD = ModPoly

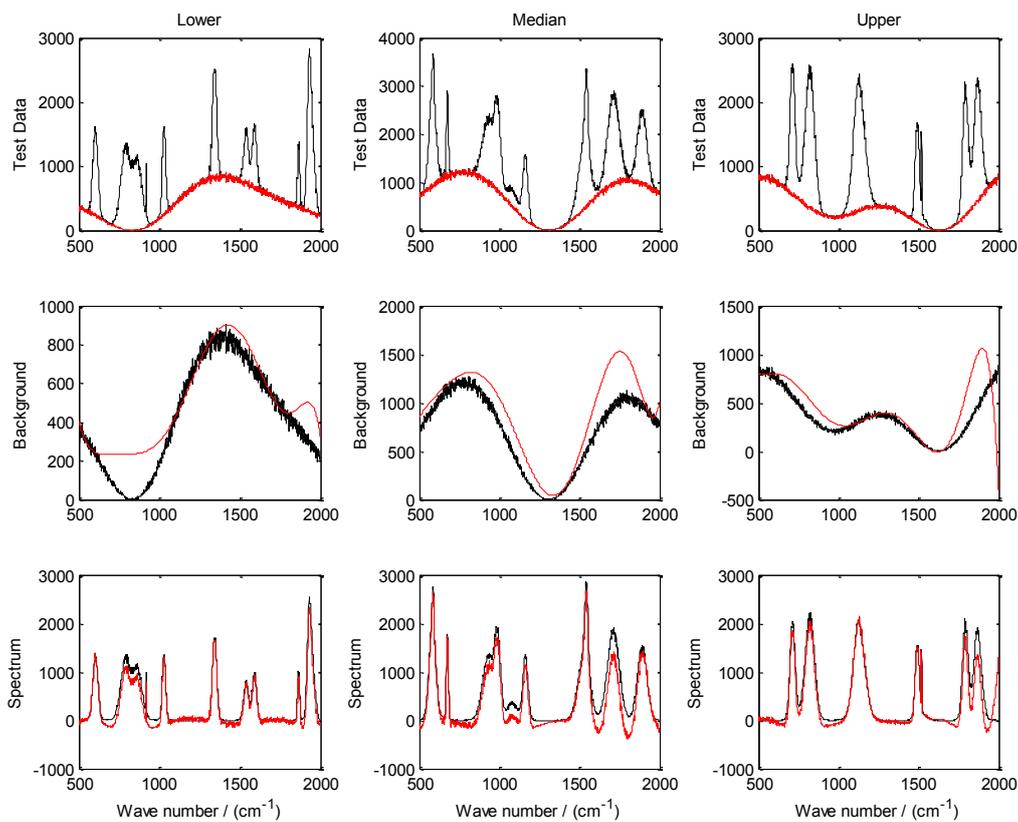

Figure S27. Np = 12; $P_1$ = 0; $P_2$ = 2000; $A_1$ = 0; $A_2$ = 400; Sf = 1 (BNR = 15); METHOD = IModPoly

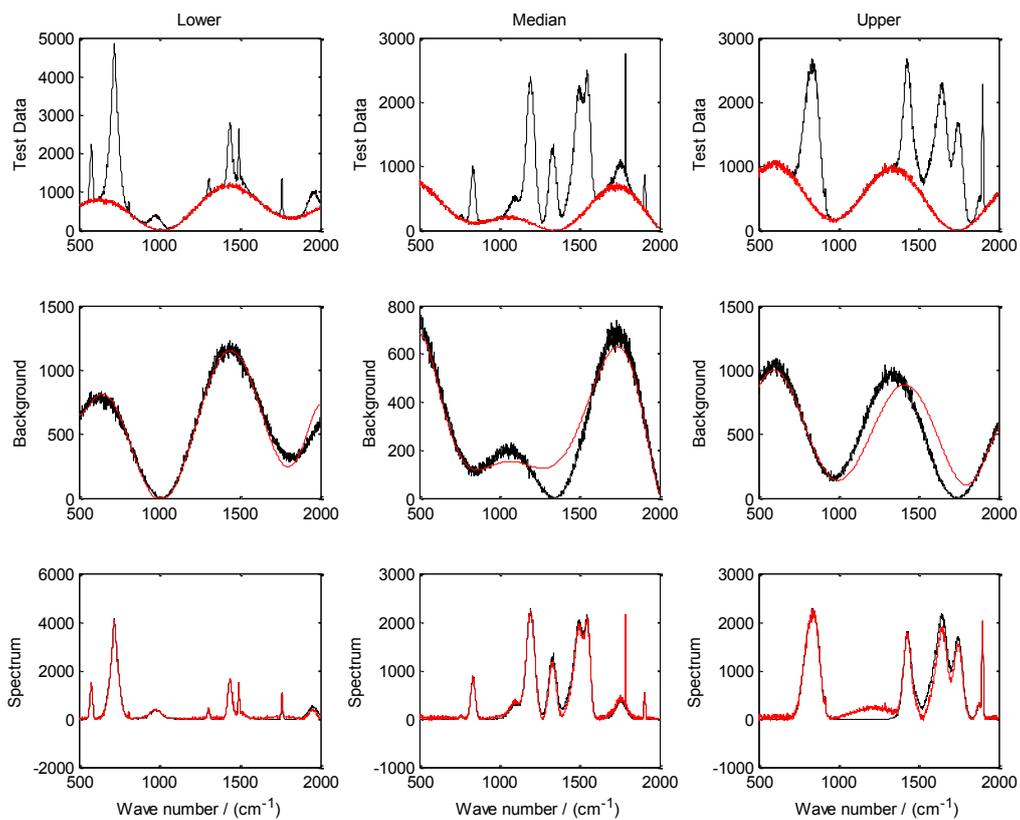

Figure S28.  Np = 12; $P_1$ = 0; $P_2$ = 2000; $A_1$ = 0; $A_2$ = 400; Sf = 1 (BNR = 15); METHOD = APoly

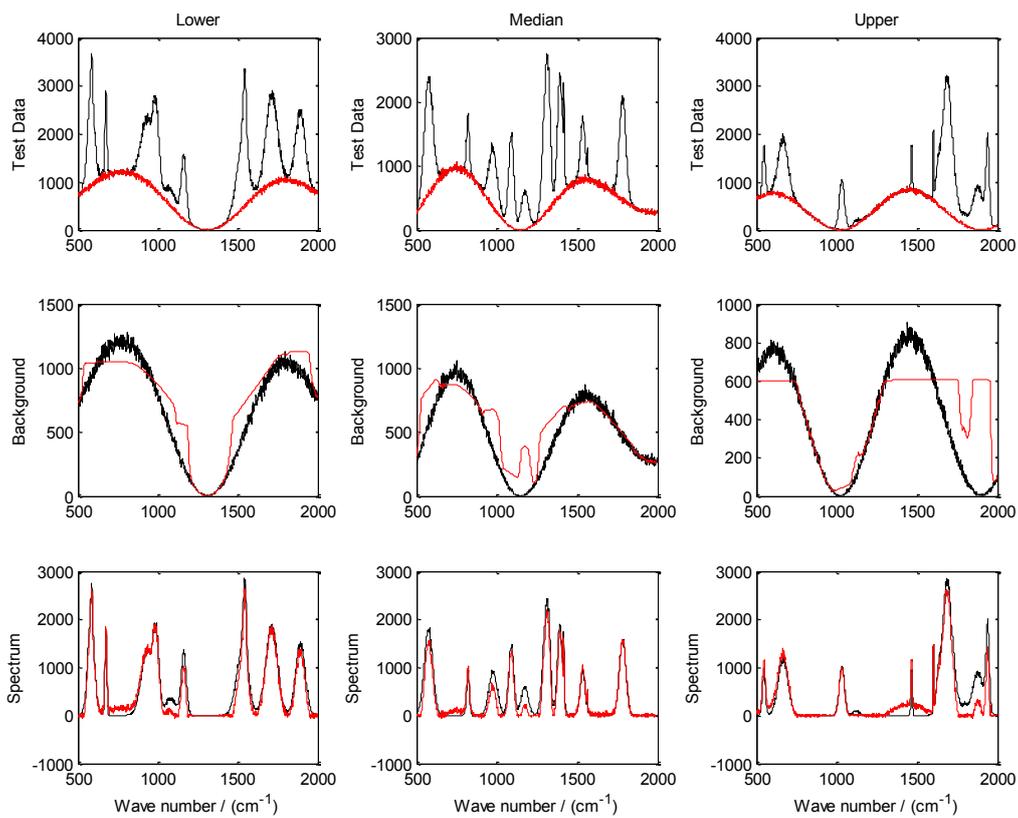

Figure S29.   Np = 12; $P_1$ = 0; $P_2$ = 2000; $A_1$ = 0; $A_2$ = 400; Sf = 1 (BNR = 15); METHOD = WPLS

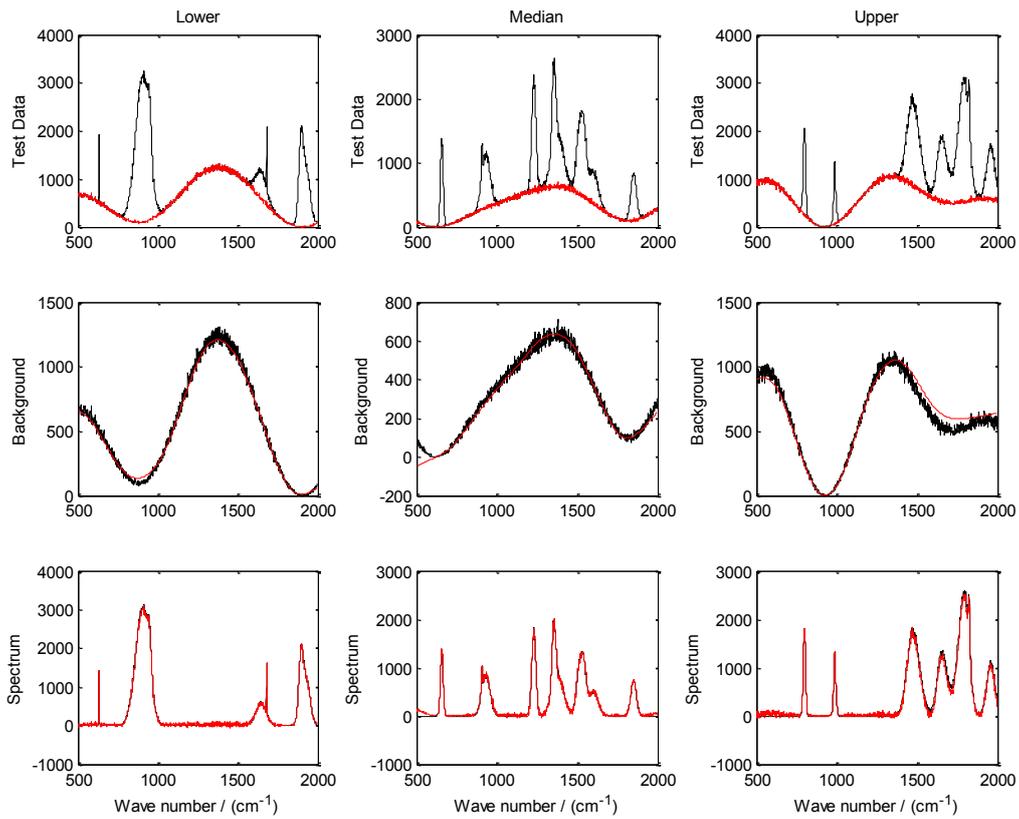

Figure S30.  Np = 12; $P_1$ = 0; $P_2$ = 2000; $A_1$ = 0; $A_2$ = 400; Sf = 1 (BNR = 15); METHOD = APLS.

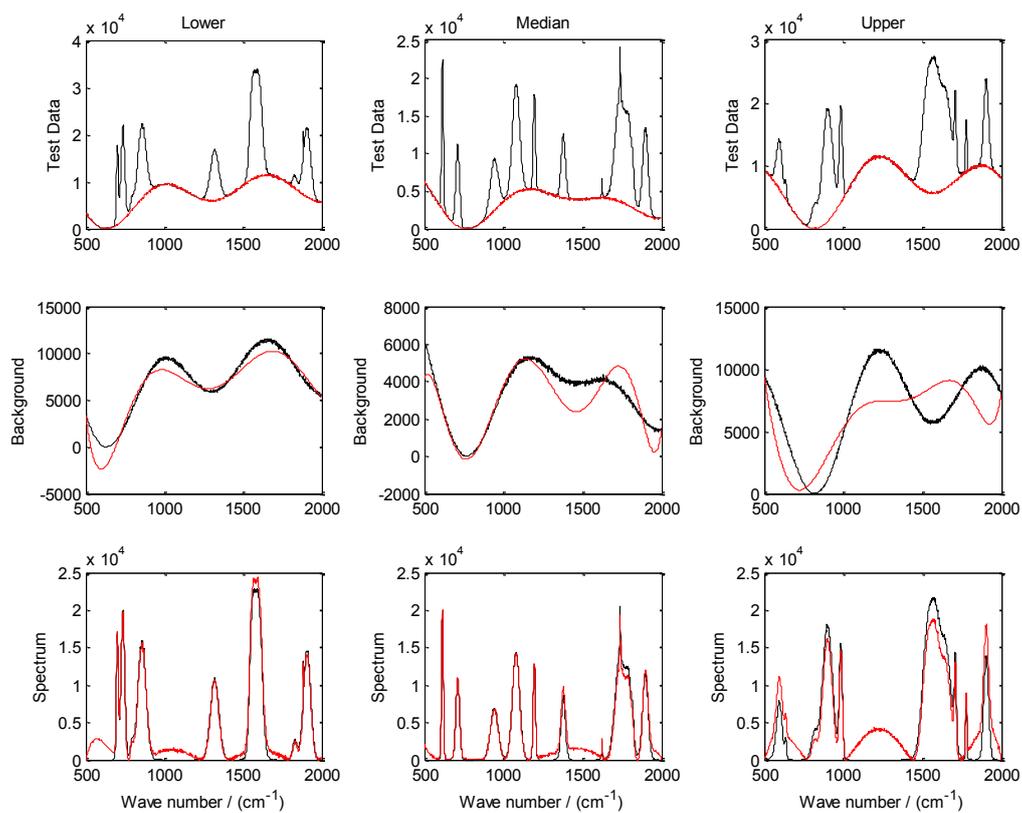

Figure S31. Np = 12; $P_1$ = 0; $P_2$ = 2000; $A_1$ = 0; $A_2$ = 400; Sf = 10 (BNR = 46.3); METHOD = ModPoly.

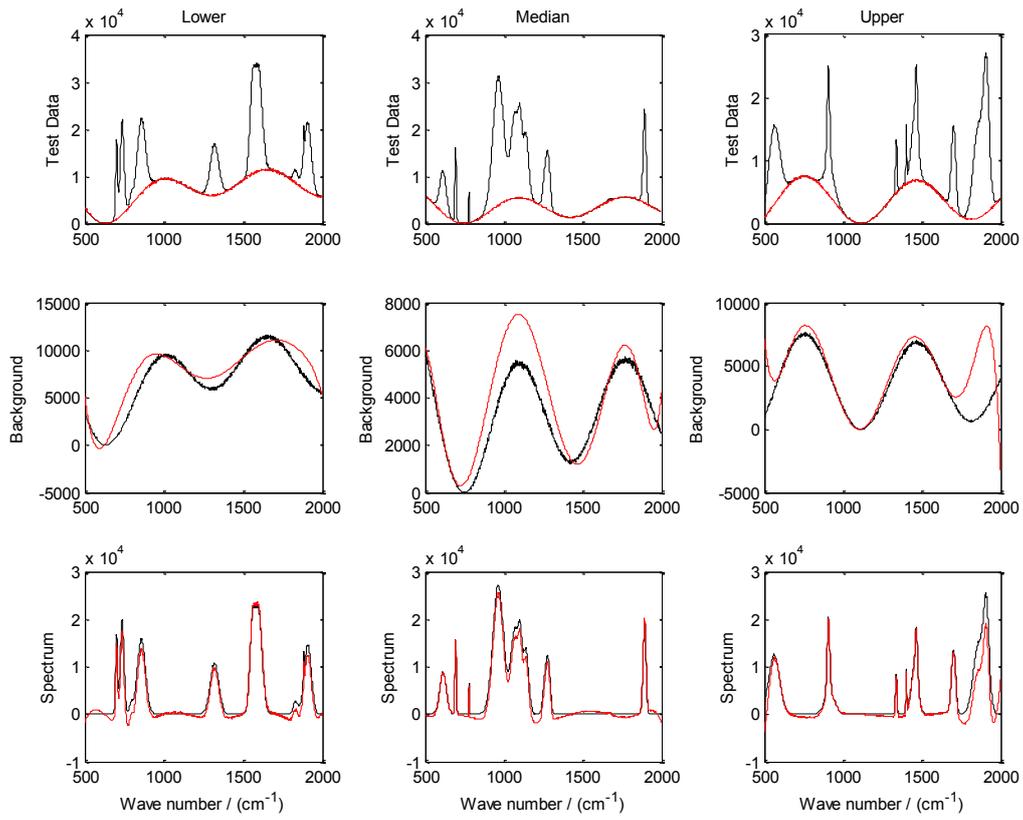

Figure S32.  Np = 12; $P_1$ = 0; $P_2$ = 2000; $A_1$ = 0; $A_2$ = 400; Sf = 10 (BNR = 46.3); METHOD = IModPoly.

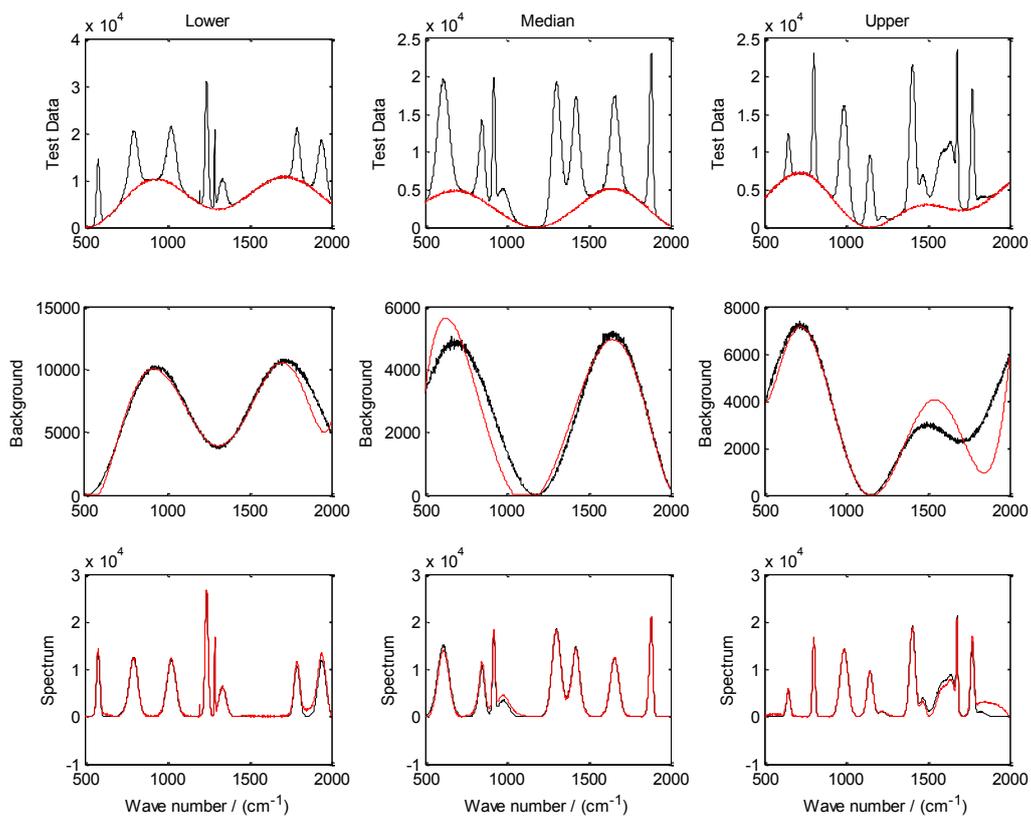

Figure S33.   Np = 12; $P_1$ = 0; $P_2$ = 2000; $A_1$ = 0; $A_2$ = 400; Sf = 10 (BNR = 46.3); METHOD = APoly.

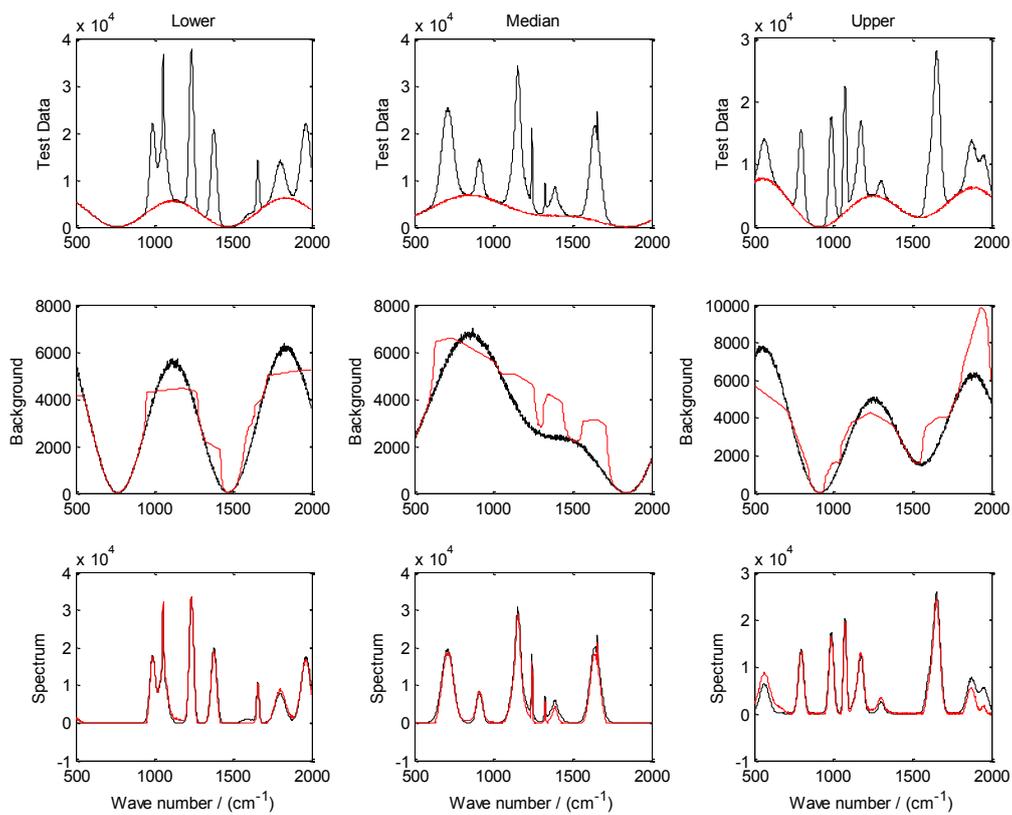

Figure S34.  Np = 12; $P_1$ = 0; $P_2$ = 2000; $A_1$ = 0; $A_2$ = 400; Sf = 10 (BNR = 46.3); METHOD = WPLS.

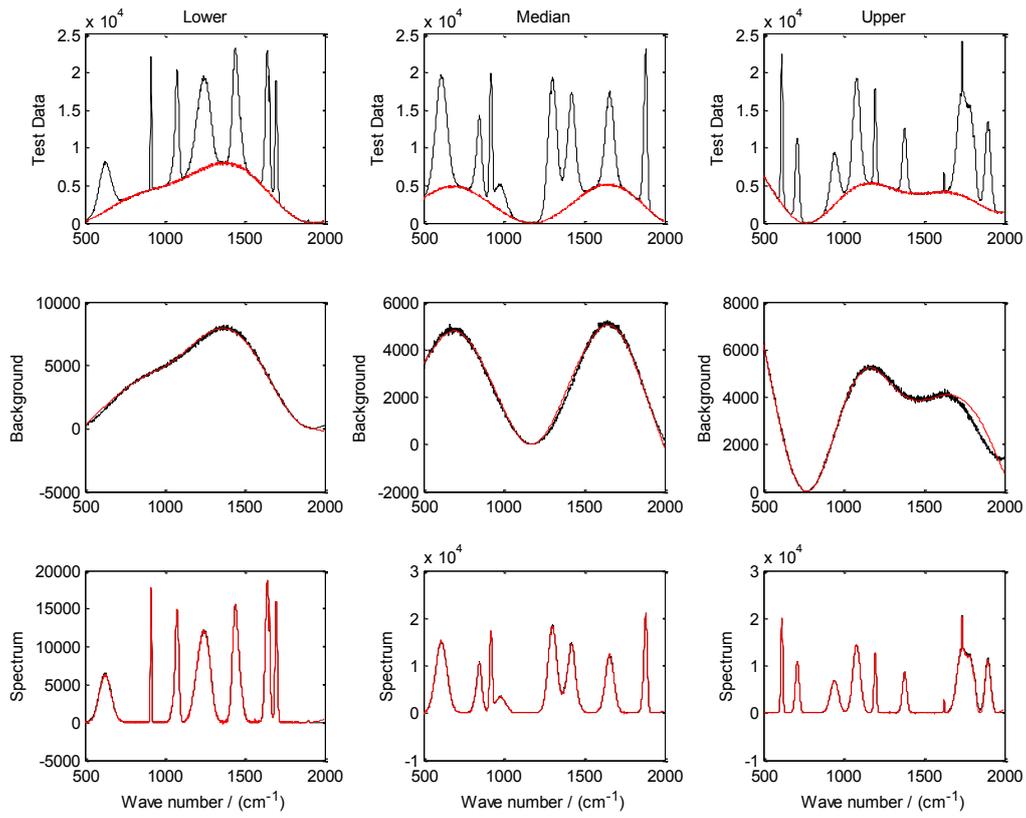

Figure S35. Np = 12; $P_1$ = 0; $P_2$ = 2000; $A_1$ = 0; $A_2$ = 400; Sf = 10 (BNR = 46.3); METHOD = APLS.